\documentclass[]{iopart}
\usepackage{iopams}
\usepackage{graphicx}
\usepackage{bbm,bm,epsfig}

\newcommand{\be}{\begin{eqnarray}}
\newcommand{\ee}{\end{eqnarray}}
\newcommand{\eins}{\mbox{$1 \hspace{-1.0mm}  {\bf l}$}}
\newcommand{\ket}[1]{\left|{#1}\right\rangle}

\newcommand{\one}{\mbox{$1 \hspace{-1.0mm}  {\bf l}$}}  
\newcommand{\C}{{\mathbb{C}}}

\begin{document}



\topical[Entanglement purification and QEC]{Entanglement purification and quantum error correction}

\author{W. D\"ur$^{1,2}$ and H. J. Briegel$^{1,2}$}

\address{$^1$ Institut f{\"u}r Theoretische Physik, Universit{\"a}t Innsbruck,
Technikerstra{\ss}e 25, A-6020 Innsbruck, Austria\\
$^2$ Institut f\"ur Quantenoptik und Quanteninformation der \"Osterreichischen Akademie der Wissenschaften, Innsbruck, Austria.}

\date{\today}

\begin{abstract}
We give a review on entanglement purification for bipartite and multipartite quantum states, with the main focus on theoretical work carried out by our group in the last couple of years. We discuss entanglement purification in the context of quantum communication, where we emphasize its close relation to quantum error correction. Various bipartite and multipartite entanglement purification protocols are discussed, and their performance under idealized and realistic conditions is studied. Several applications of entanglement purification in quantum communication and computation are presented, which highlights the fact that entanglement purification is a fundamental tool in quantum information processing. 
\end{abstract}


\maketitle

\tableofcontents


\section{Introduction}

Entanglement is a unique phenomenon of quantum mechanics that has puzzled generations of physicist. While initially the discussion was mainly driven by conceptual and philosophical considerations, e.g. regarding hidden variable theories, in recent years the focus has shifted to practical aspects and applications. It was realized that entanglement constitutes a valuable resource that can be used for various tasks in quantum information processing, many of which have in the meantime been experimentally demonstrated with a number of systems including nuclear magnetic resonance, photons, light beams, ions, neutral atoms, atomic ensembles, cavity QED, Josephson junctions and quantum dots. The remarkable experimental progress is even exceeded by the vast theoretical achievements that have lead to a new theory of quantum information, with quantum cryptography and quantum computation as most prominent offsprings. 

In this context, the generation and maintenance of high--fidelity entanglement is a central problem. In the last decade, the controlled manipulation of certain systems in such a way that entangled states can be produced on demand has become possible. 
However, noise in such control operations as well as interactions with an uncontrollable environment have the effect that the desired entangled states are produced only with a certain non--unit fidelity. 
Similarly, entangled states that are ground states of certain (strongly coupled) systems are not obtainable in the lab due to thermal fluctuations that lead to a thermal (mixed) state with non--unit fidelity for any non--zero temperature.  Several ways to protect quantum information in general and entangled quantum states in particular have been designed in recent years. These methods include quantum error correction, where quantum information is protected by using a certain encoding, quantum error avoiding schemes as well as entanglement purification. The latter is the main subject of this review article. 

In entanglement purification \cite{Be96a,Be96,De96}, several copies of noisy, non--maximally entangled states are manipulated in such a way that a fewer number of copies with a reduced amount of noise are produced. The entanglement of the total ensemble is concentrated or distilled in a few copies, which hence contain a larger amount of entanglement and have higher fidelity with respect to a maximally entangled states. We will use the term {\em entanglement distillation} to refer to the manipulation of an ensemble of states in such a way that (a reduced number of) maximally entangled states are distilled. Entanglement distillation uses entanglement purification as a building block to increase the information about the ensemble, and hence to achieve this aim. 
The purified states can then eventually be used for various entanglement--based applications, ranging from quantum teleportation to quantum computation.
It is worth pointing out that entanglement purification has been experimentally demonstrated for single photons \cite{Pa01,Kw01,Pa04,Ya03,Re05}, and very recently also for atoms \cite{Re06}.

Entanglement purification was originally introduced in the context of quantum communication as a solution to the problem of communication over noisy quantum channels \cite{Be96a,Be96,De96}. We will discuss this aspect of entanglement purification in Sec. \ref{ECC}, where we also briefly review quantum error correction as an alternative solution. Sec. \ref{Protocols} includes an overview and a detailed description of known bipartite entanglement purification protocols, while Sec. \ref{relation} establishes the close connection between quantum error correction and entanglement purification. The influence of noise in (local) control operations is discussed in Sec. \ref{realistic}, while Sec. \ref{applications1} illustrates a number of applications of entanglement purification protocols, including long distance quantum communication and quantum computation. We then move to multipartite systems, where we describe multipartite entanglement purification protocols and the role of noise in these protocols in Sec. \ref{multipartiteEPP},  and illustrate possible applications in Sec. \ref{applications2}. We provide an outlook on future perspectives of entanglement purification, including a number of relevant open problems, in Sec. \ref{outlook}. 

We remark that in this review article we only touch upon the (extensive) subject of quantum error correction. For a more comprehensive treatment of quantum error correction we refer the reader for example to Refs. \cite{reviewQEC,St98,Go97} (see also \cite{Be96,St96,QECC1,Ma96,Ca96,QECC2,QECC3,QECC4,Pr98,OQEC}). We concentrate on theoretical aspects of entanglement purification and its applications, where our main focus lies on the work which has been carried out by our group in the last couple of years.


\section{Quantum communication via noisy channels}\label{ECC}

Quantum cryptography is the most advanced application of quantum information processing, where even first commercial systems can be purchased and are probably already used in practice. The transmission of quantum information over noisy channels in such a way that their quantum nature is sufficiently preserved is the central problem in this context. As quantum information is unavoidably affected by noise, ways to protect it need to be designed in order to ensure (almost) error free, i.e. noiseless, exchange of quantum information.  

Even though quantum information cannot be cloned perfectly or amplified without changing their quantum nature, different techniques are known to protect quantum information from noise. The two solutions to the problem are provided by
\begin{itemize}
\item[(i)] quantum error correction 
\item[(ii)] teleportation in combination with entanglement purification. 
\end{itemize}
In (i), quantum information is encoded in a larger dimensional Hilbert space and in this way protected from noise. In the second approach (ii), the problem of transmitting unknown quantum information over a noisy channel is replaced by the task of distributing a known entangled state with sufficiently high fidelity, which can then be used, by means of teleportation, to transmit arbitrary quantum information. Entanglement purification is thereby the tool to obtain (known) high--fidelity entangled pairs. 

We first specify the set--up we have in mind. We consider two spatially separated parties $A$ (Alice) and $B$ (Bob) who wish to communicate with each other, i.e. transmit quantum information. They are connected via a possibly noisy quantum channel described by a completely positive map ${\cal E}$,  and in addition by a classical channel which either only allows for classical communication from $A \to B$ (one way classical communication), or for classical communication between $A \to B$ and $B \to A$ (two--way classical communication). In addition, we assume that Alice and Bob can locally manipulate their quantum states and have access and control of auxiliary systems. This set of local operations and classical communication is denoted by LOCC. 
For qubits, a general quantum channel ${\cal E}$ can be written as 
\be\label{channel}
{\cal E}\rho = \sum_{k,l=0}^{3} p_{k,l} \sigma_k \rho \sigma_l,
\ee
where $\sigma_j$ denote Pauli operators with $\sigma_0 =\one, \sigma_1=\sigma_x, \sigma_2=\sigma_y, \sigma_3= \sigma_z$. Often, we will consider Pauli-diagonal channels ${\cal E_P}$ which are of the form
\be\label{Paulidiag}
{\cal E_P}\rho = \sum_{k=0}^{3} p_{k} \sigma_k \rho \sigma_k.
\ee
Notice that any quantum channel ${\cal E}$ can be brought to Pauli--diagonal form by means of depolarization (see Ref. \cite{Du05SF}) in such a way that the diagonal elements are not altered, $p_k=p_{k,k}$. This often allows one to restrict considerations to Pauli--diagonal channels, and makes such channels particularly important. A special instance of a Pauli--diagonal channel is the depolarizing (or white noise) channel, where $p_1=p_2=p_3 = (1-p_0)/3$, which is described by a single parameter $p=p_0$.

We will discuss the basic principles of quantum error correction and entanglement purification in the following.

\subsection{Quantum error correction}

In standard quantum error correction (QEC) (see e.g. Refs. \cite{reviewQEC,St98,Go97} for reviews, and also \cite{Be96,St96,QECC1,Ma96,Ca96,QECC2,QECC3,QECC4,Pr98,OQEC}), quantum information is protected by encoding one logical qubit of quantum information into several physical qubits, or more generally $k$ logical qubits into $n$ physical qubits. The basic idea of such a redundant encoding is borrowed from classical coding and error correction, although additional requirements have to be met for quantum error correction, e.g. the preservation of arbitrary superposition states. In the simplest case where one qubit is encoded into $n$ qubits, we define logical qubits $|0_L\rangle, |1_L\rangle$ as two orthogonal states $|\psi_0\rangle=|0_L\rangle, |\psi_1\rangle=|1_L\rangle \in \C^{2^n}$. Any (unknown) qubit in a state $\alpha |0\rangle + \beta |1\rangle$ is encoded via a unitary encoding operation $U_E$ $\in SU(2^n)$ yielding an encoded state
\be
(\alpha|0\rangle + \beta |1\rangle) \otimes |0\rangle^{\otimes n-1} \rightarrow (\alpha|0_L\rangle + \beta |1_L\rangle).
\ee  
The choice of $|\psi_0\rangle, |\psi_1\rangle$ is crucial for the error correcting properties of the code. The basic idea is to use states  $|\psi_0\rangle, |\psi_1\rangle$ where the two--dimensional subspace $S$ spanned by $\{|\psi_0\rangle, |\psi_1\rangle\}$ is mapped to orthogonal two--dimensional subspaces by error operators. By measuring appropriate two--dimensional projectors, one can distinguish between these subspaces. Importantly, coherent superpositions within each of the subspaces are not altered and hence quantum information is preserved. Notice that these measurements also discretize errors, i.e. an independent treatment of error operators is justified. Consider as an example a $n$--qubit code that is capable of correction an arbitrary single qubit error, specified by a single--qubit Pauli error $\sigma_i^{(a)}$, acting on one of the $n$ qubits. If the action of any of these Pauli operators $\sigma_i^{(a)}$ is such that the two--dimensional subspace $S$ is transformed to a subspace $S_{i,a}$, and all these subspaces are pairwise orthogonal, then the corresponding code is capable of correcting for all such errors. The minimum number $n$ of qubits for which this is possible is given by 5, as there are $3n$ different error operators, and the possible number of orthogonal two--dimensional subspaces is given by $2^{n-1}$. In fact, error correction codes that can protect a single qubit from an arbitrary single qubit error using only 5 qubits are known \cite{Be96,QECC3}. In a similar way, one can construct quantum error correcting codes where $k$ logical qubits are encoded into $n$ physical qubits, and where a total of $m$ errors can be corrected. The basic idea of such a construction is still that each correctable error operator should map the relevant information carrying subspace $S$ of dimension $2^k$ to an orthogonal subspace $S_j$, and all these subspaces are pairwise orthogonal and hence distinguishable. In this way the type of error can be detected and corrected by determining the correspond $j$ via $2^k$--dimensional projective measurements. In addition, quantum information stored in coherent superpositions of states within the subspace $S$ remain unaltered.

Codes can also be designed to only {\em detect} errors. For error detection, it is only required that an error operator $O_j$ maps states within the subspace $S$ to the orthogonal subspace $S^\perp$, however it is no longer necessary that different error operators map all states within the subspace $S$ to pairwise orthogonal subspaces $S_j$. This implies that different error operators can lead to the same output state, and hence are indistinguishable. Nevertheless, the occurrence of an error can still be determined. Notice that the possibility to detect certain kinds of e.g. multi--qubit errors in addition to the possibility to distinguish between any kind of single--qubit errors is often an additional feature of error correcting codes.
Error correcting codes can also be applied in a concatenated fashion, that is an encoded qubit is once more encoded on a next level using now encoded qubits $\alpha|0_L\rangle + \beta |1_L\rangle$ of level one as basic elements. Such concatenated error correction codes lie at the heart of a fault--tolerant implementation of quantum computation, but are also discussed in the context of long--range quantum communication using QEC \cite{Kn96}. 

A general class of quantum error correcting codes of particular importance are the Calderbank-Shor-Steane (CSS) codes \cite{Ca96,St96} where encoding and decoding circuits consist of Clifford operations only \cite{Go97}. We will mainly consider such codes here. These codes belong to the class of {\em stabilizer codes}, which allows for a simplified description and error analysis. We also mention that recently the concept of operator quantum error correction \cite{OQEC}, as well as topological protection of quantum information \cite{Ki03} have been introduced. 

\subsubsection{Long--distance quantum communication using QEC}

For long--range quantum communication, one protects an (unknown) qubit by encoding it with a concatenated quantum error correcting code \cite{Kn96}. The encoded quantum information is then sent through a noisy channel over a short distance, where the distance is chosen such that the probability for an error on the logical qubit is sufficiently small. More precisely, the error probability must be below a certain threshold such that error correction is still possible. Then an error correction step is performed, which involves decoding and measurements or direct error syndrome extraction, as well as correction. The signal is again encoded and sent further through a small segment of the channel. Concatenation of the error correction code ensures that errors on physical qubits below a certain threshold become exponentially suppressed at higher concatenation levels. This yields to a perfect transmission of quantum information with only polynomial overhead in additional qubits. To guarantee a fault--tolerant transmission, errors in coding and decoding operations at the error correction stations need to be taken into account. Although this is possible, one obtains a rather stringent error threshold of the order of $10^{-4}$ for local control operations, and also a small amount of tolerable channel noise at the order of percent \cite{Kn96}, significantly restricting the length of the segments.

\subsubsection{Channel capacities and capacities of QEC codes}

The capacity of a quantum error correcting code (QECC) is defined as the maximum asymptotic rate of reliable transmission of unknown quantum information through a noisy quantum channel, using the QECC to encode the states before transmission and decode them afterwards. More precisely, we consider a quantum channel ${\cal E}$ which is described by a completely positive 
trace preserving linear map ${\cal E}$ from the input Hilbert space ${\cal H}_c$ 
to the output Hilbert space ${\cal H}_o$. A quantum error correcting code is associated to coding (${\cal C}$) and decoding (${\cal D}$) operations. The coding operation ${\cal C}$ maps an input state $|\phi\rangle \in {\cal H}_{\rm in}$ to an encoded state of $n$ systems in Hilbert space ${\cal H}_c^{\otimes n}$, which are then transmitted through a noisy quantum channel ${\cal E}^{\otimes n}$ and decoded afterwards. These coding and decoding operations define a $(n,\epsilon)$ code if one achieves transmission with fidelity larger than $1-\epsilon$ for all possible system states, 
\be
{\rm min}_{|\phi\rangle \in{\cal H}_{\rm in}} \langle\phi| {\cal D} 
\circ {\cal E}^{\otimes n} \circ {\cal C} (|\phi\rangle\langle\phi|) |\phi\rangle \geq 1-\epsilon.
\ee
The rate $R\equiv \log \dim {\cal H}_{\rm in}/n$ of a specific code is called achievable if for all 
$\epsilon,\delta >0$ and sufficiently large $n$ one obtains a rate $R-\delta$. 
The quantum capacity $Q(\cal E)$ of a bipartite quantum channel ${\cal E}$ is defined as the supremum of all achievable rates $R$ over all codes (see Ref. \cite{Ba98} for a rigorous definition).
 Coding and decoding operations may be assisted by forward classical communication 
($\rightarrow$) or two--way classical communication $(\leftrightarrow)$ which gives 
rise to quantum channel capacities $Q^{\rightarrow}$ [$Q^{\leftrightarrow}$] respectively. We will mainly consider $Q=Q^{\rightarrow}$ here. 
We remark that a minimal pure state fidelity $F=1-\epsilon$ for all $|\phi\rangle \in 
{\cal H}_{\rm in}$ implies an entanglement fidelity $F_e \geq 1- 3\epsilon/2$ for all 
density operators $\rho$ whose support lies entirely in that subspace \cite{Ba98}. 
That is, when transmitting part of an entangled state $|\Phi\rangle$ which is a 
purification of $\rho$, the resulting state has fidelity $F\geq 1-3 
\epsilon/2$ with respect to $|\Phi\rangle$.

\subsection{Quantum communication via noisy entanglement purification and teleportation}

An alternative to direct transmission of quantum information over a (noisy) channel is provided by noisy entanglement purification followed by teleportation. More precisely, as shown in Ref. \cite{Be93}, a maximally entangled pair 
\be
|\phi^+\rangle = \frac{1}{\sqrt{2}}(|0\rangle_A|0\rangle_B + |1\rangle_A|1\rangle_B)
\ee
shared between $A$ and $B$ can be used as a resource to transmit one qubit of information from Alice to Bob, using only local resources and classical communication of two classical bits. In this sense, a maximally entangled pair serves as a perfect quantum channel. Hence the problem of transmitting unknown quantum information is shifted to the problem of generating a maximally entangled state. 

We start by briefly reviewing the teleportation protocol \cite{Be93}, while entanglement purification is treated in more detail below. The teleportation protocol consists of the following steps: 
\begin{itemize}
\item[(i)] Alice performs a local Bell measurement in the basis $\{|\phi_{j}\rangle_{A'A}\}$ with $|\phi_j\rangle = \eins \otimes \sigma_j |\phi^+\rangle$ on qubit $A'$ to be teleported and qubit $A$ of the maximally entangled state; 
\item[(ii)] Bob performs a correction operation $\sigma_j$ on qubit $B$ depending on the outcome $j$ of the measurement. 
\end{itemize}
Step (ii) involves classical communication from Alice to Bob. Please note that in this way not only the local quantum information stored in the qubit $A'$ is transferred to $B$, but the qubit $B$ takes over completely the role of qubit $A'$, in particular also its entanglement with additional particles. The latter property can be used for entanglement swapping \cite{Zuxx1,Zuxx2,Zuxx3,Zuxx4}, where entanglement between systems $A C_1$ and $C_2 B$ leads to an entangled state between systems $AB$ by teleporting $C_1$ to $B$. 
Teleportation has been experimentally demonstrated for single photons \cite{Tphoton1,Tphoton2,Tphoton3}, light beams \cite{Tlight} and atoms \cite{Tatom,Tatom1}, and very recently also to for the transmission of quantum information between different media, namely from light to matter \cite{Tlighttoatom}.

For quantum communication over noisy channels, teleportation alone is not sufficient. When sending one qubit of a maximally entangled pair through a noisy channel, one ends up with a noisy, non--maximally entangled state. Although such a noisy state can still be used for quantum teleportation, the fidelity of the teleported qubit is reduced. However, one can in principle produce many copies of the noisy entangled pairs and purify these pairs using entanglement purification, i.e. increase the entanglement of a few copies. This is possible since the desired state is {\em known}, in contrast to the general situation in direct quantum communication where the states may be unknown.

Entanglement purification is a fundamental tool in quantum information processing. We will concentrate in this section on its application in quantum communication, where entanglement purification together with teleportation provides a scheme for error--free quantum communication over noisy channels. The basic idea is to create several copies of noisy entangled states, e.g. by sending parts of locally created entangled pairs through noisy quantum channels. These states are then processed locally, more precisely a sequence of local operations (assisted by one or two-way classical communication) is applied in such a way that a reduced number of pairs with an increased fidelity is generated. Iteration of such protocols eventually leads to maximally entangled states. The entanglement is purified at the cost of obtaining a smaller number of copies. This can be done in a systematic way, and several entanglement purification protocols are known which can achieve this task. The purification protocols can be grouped into distillation protocols, and recurrence and pumping schemes. In distillation protocols, an ensemble of many (identical) copies is manipulated and a few pairs with improved fidelity are generated. In recurrence and pumping schemes, a certain elementary purification step is repeated several times, resulting in pairs with improved fidelity. We will discuss such entanglement purification protocols in the following section in more detail.

\subsubsection{One--way and two--way classical communication}

The protocols differ in the number of initial and final copies of the states, and the allowed additional resources, most importantly one--way or two--way classical communication. 

A protocol may operate on $N$ copies of noisy entangled states and produce $M \leq N$ purified copies as output. In case of one--way classical communication, one measures $N-M$ copies and uses the obtained information to choose a proper correction operation on the remaining pairs. Notice that for the choice of these correction operations, Alice only has access to the local measurement outcomes in $A$, while Bob has access to outcomes of measurements in $A$ and $B$. In particular, this implies that the parties {\em cannot} decide to discard certain pairs based on joint measurement outcomes, as it is used in several entanglement purification protocols. To make this possible, one needs two--way classical communication. Protocols based on one--way classical communication turn out to be equivalent to quantum error correction \cite{Be96} in a sense we will specify below, and one may say that entanglement purification runs in {\em error correction mode}. On the other hand, protocols with two--way classical communication make use of additional resources and are provably superior to quantum error correction and one--way entanglement purification \cite{Be96,AsPhD}. Such two--way protocols can also run in an {\em error detection mode}. 

\subsubsection{Purification range and yield of a protocol}

The yield of a protocol with respect to a certain state is a central quantity that determines how efficient a protocol is. Consider $N$ copies of a mixed state, $\rho^{\otimes N}$, which are processed by a certain entanglement purification protocol ${\cal P}$ that may be applied in an recursive way. After this procedure, $M$ (exact) copies of a maximally entangled state are obtained. The yield of the protocol ${\cal P}$ with respect to the state $\rho$ is defined as
\be
Y_{\rho,{\cal P}} = \frac{M}{N},
\ee 
i.e. the ratio of number of maximally entangled states obtained by the protocol over the total number of initial copies of the state $\rho$, where often the limit $N \to \infty$ is considered. More precisely, one considers an entanglement purification protocol and the corresponding LOCC transformation $\rho^{\otimes N} \rightarrow \tilde \Gamma_N$, where we are interested only in the reduced density operator of $M$ copies, $\Gamma_M = {\rm tr}_{M-1,\ldots N} \tilde \Gamma_N$. We demand that for all $\epsilon >0$, the fidelity of $\Gamma_M$ with respect to $M$ copies of a maximally entangled state $|\Phi\rangle = |\phi^+\rangle^{\otimes M}$ should be larger than $1-\epsilon$, $F=\langle\Phi|\Gamma_M|\Phi\rangle \geq 1-\epsilon$. The yield is then determined by the ratio $M/N$ of the maximum $M$ for which this is the case, in the asymptotic limit of $N \to \infty$. The maximum of the yield over {\em all} protocols (or equivalently all sequences of local operations and classical communications (LOCC)) is also called the distillable entanglement $D_\rho$, and again one may consider these quantities with respect to one way or two--way classical communication, $Y_{\rho,{\cal P}}^\rightarrow, Y_{\rho,{\cal P}}^\leftrightarrow, D_\rho^\rightarrow, D_\rho^\leftrightarrow$. The distillability problem, i.e. the question whether there exists a LOCC protocol that can generate maximally entangled states from (infinitely) many copies of a state, has been extensively studied in recent years, however a complete solution has not been obtained so far. What is, however, known are {\em necessary} conditions for distillability (e.g. that the partial transposition of the density operator is non--positive \cite{Ho97}), as well as {\em sufficient} criteria. In particular, any entanglement purification protocol provides a sufficient criterion for distillability.

We remark that such a strict definition of yield actually implies that many entanglement purification protocols have zero yield, although they can produce entangled states with arbitrary high fidelity. We will consider later a modified definition of the yield, where the condition of arbitrary accuracy for the produced states is dropped and replaced by a condition for a certain {\em fixed} fidelity $F=1-\epsilon_0$. Notice that whenever one considers noise in local control operation, perfect entanglement purification is impossible, i.e. no entanglement purification protocol is capable of producing perfect maximally entangled states with fidelity $F=1$. 

The purification range (or basin) of a protocol is defined as the set of all input states $\rho$ that can be purified by the protocol, i.e. where maximally entangled states can be generated from $N$ copies of $\rho$ in the limit of $N\to \infty$. Often, certain families of density operators specified by a single or a few parameters are considered. An example of such a family are the so--called Werner states (see Eq. (\ref{Wernerstates}) below) which are mixtures of a maximally entangled state and a completely mixed state. In this case, the purification range can be obtained analytically, yielding a simple condition that can be expressed in terms of the fidelity of the state. In general a complete characterization of the purification basin for a given protocol is very complicated. However, bounds on the purification range can be established which are based on either necessary conditions for distillation, or on sufficient conditions for entanglement purification using a specific protocol (e.g. that the fidelity with respect to a maximally entangled state be above a certain threshold value).


\section{Basic bipartite entanglement purification protocols}\label{Protocols}

We now turn to explicit entanglement purification protocols. A number of different protocols exist, which differ in their purification basin (i.e. the set of states they can purify), the efficiency, and the number of copies of the states they operate on. In the following we will consider filtering protocols (which operate on a single copy), recurrence protocols (which operate on two copies simultaneously at each step) as well as hashing and breeding protocols (which operate simultaneously on a large number $N \to \infty$ of copies). We also discuss (intermediate) $N \to M$ protocols, which operate on $N$ input copies and produce $M$ output copies. The latter protocols can also be run in a recursive way.

\subsection{Filtering protocol}
The simplest protocols operate on a single copy of the mixed state $\rho$ and consist in the application of local filtering measurements (including weak measurements). A weak measurement may e.g. be realized by a joint, local operation on the system and a (possibly high dimensional) ancilla, followed by a von Neumann measurement of the ancilla. Hence sequences of local operations, including weak measurements, are applied in such a way that for specific measurement outcomes the resulting state $\sigma$ is more entangled than the initial state $\rho$. Note that the output state $\sigma$ is obtained only with a probability $p < 1$. Mixed states where such a filtering method can be applied include certain rank two states 
of the form \cite{Gi98}
\begin{eqnarray}
\rho=F |\Psi^+\rangle\langle \Psi^+| + (1-F) |00\rangle\langle00|,
\end{eqnarray}
where $|\Psi^+\rangle=1/\sqrt{2}(|01\rangle+|10\rangle)$. Application of the local operators $O_A= O_B =\sqrt{\epsilon} |0\rangle\langle 0| + |1\rangle\langle 1|$ (which correspond to a specific branch of a local positive operator valued measure, POVM) lead to a non--normalized state of the form 
\be
\rho'= F \epsilon |\Psi^+\rangle\langle \Psi^+| + (1-F) \epsilon^2 |00\rangle\langle 00|. 
\ee
The fidelity of the resulting state is given by 
\be
F'= F\epsilon/[F\epsilon+ (1-F) \epsilon^2].
\ee
Note that for small $\epsilon$, $F' \to 1$, that is states arbitrarily close to the maximally entangled state $|\Psi^+\rangle$ can be created. However, the probability to obtain the desired outcome, $p_{\rm suc}=F\epsilon+ (1-F) \epsilon^2$, goes to zero as $\epsilon \to 0$. There is a tradeoff between the reachable fidelity of the output state and the probability of success of the procedure. The optimal filtering protocol for any mixed state of two qubits has been derived in \cite{Ve01}, and experimentally demonstrated in \cite{Wang06}.

It turns out that filtering protocols are of limited applicability for general mixed states, even for the simplest case of two qubits. In particular, as shown in Refs. \cite{Li99,Ke00}, the fidelity of a single copy of a full rank state can in general {\em not} be increase by any local operation.  This seriously restricts the applicability of filtering procedures and requires us to consider protocols on two (or more) copies of the state in order to increase the fidelity for a general class of mixed input states, and ultimately to obtain maximally entangled states.

\subsection{Recurrence protocols}

In the following we discuss a class of conceptually related protocols \cite{Be96a,Be96,De96,Du03QC} that allow one to produce states arbitrarily close to a maximally entangled pure state by iterative application. Before we go into technical details, we describe the general concept underlying these (and almost all) entanglement purification protocols. The idea of entanglement purification protocols is to decrease the degree of mixed--ness of the ensemble of mixed states. To this aim, one needs to gain information, which is done by performing suitable measurements. As the relevant information is not locally accessible, one needs to use the entanglement inherent in states of the ensemble to reveal this information. 
In fact, information about a particular sub-ensemble is obtained by first operating jointly (but still locally) on several states and then measuring one of these states.
In many protocols the remaining states are only kept if a specific measurement outcome was found. This is due to the fact that one finds for certain measurement outcomes (measurement branches) that the entanglement of the remaining states is increased, while for other outcomes it is decreased or the states are even no longer entangled. In this way it is also guaranteed that, on average, entanglement can not increase under local operations and classical communication.

\begin{figure}
\begin{center}
\includegraphics[width=0.65\columnwidth]{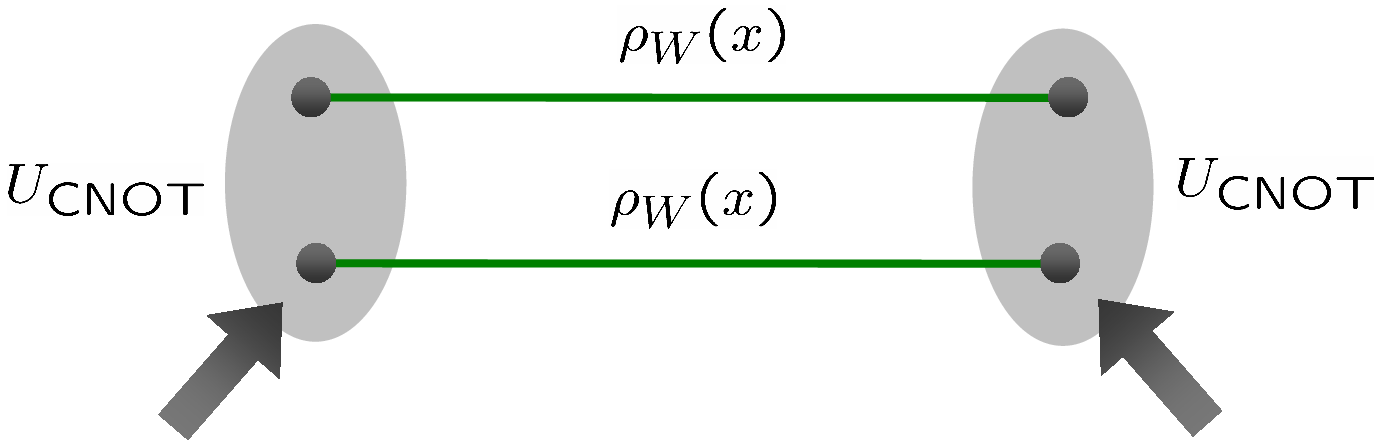}
\caption {Schematic representation of the bipartite entanglement purification.}
\label{bipartitesetupFig}
\end{center}
\end{figure}

\begin{figure}
\begin{center}
\includegraphics[width=0.85\columnwidth]{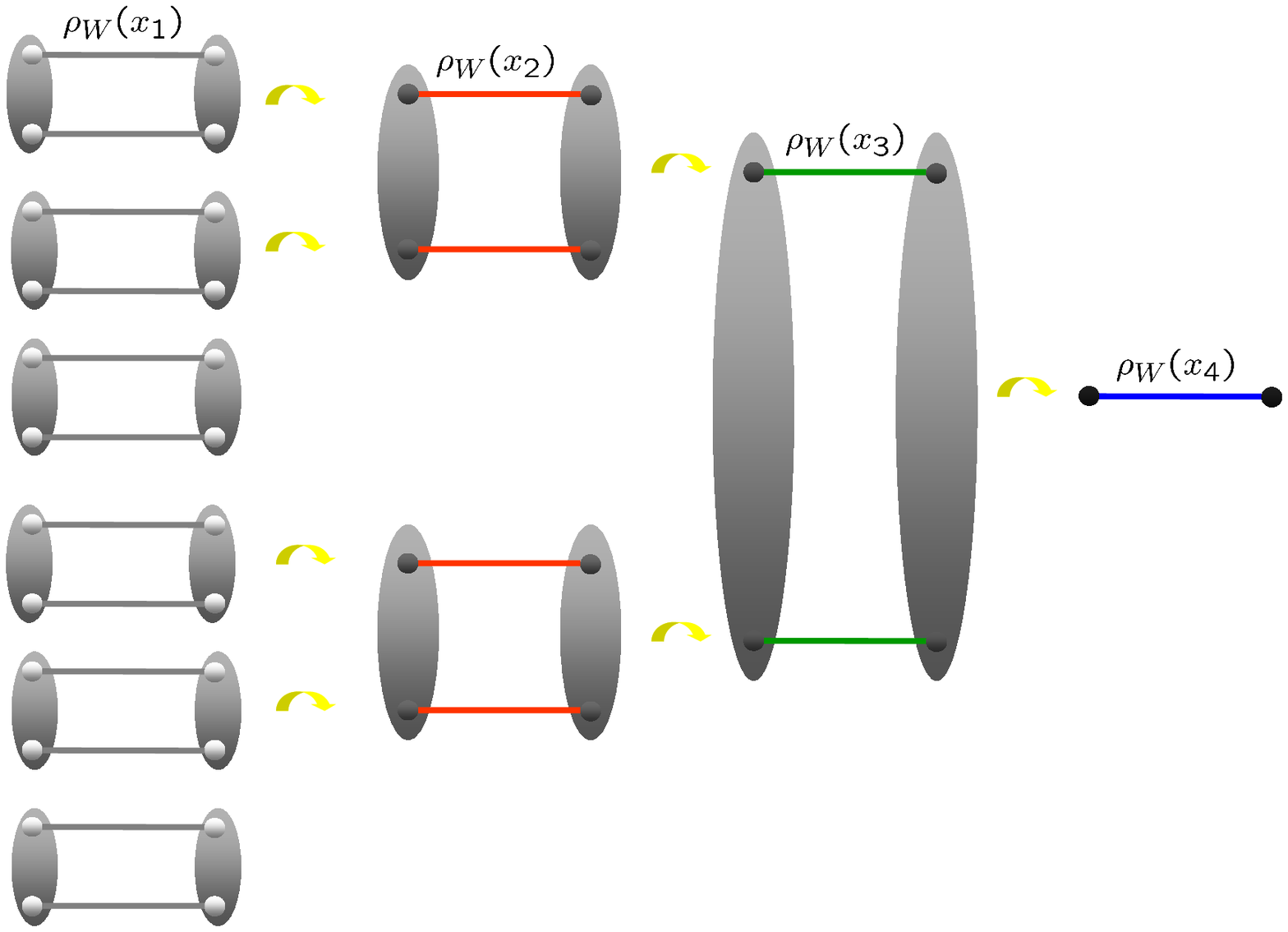}
\caption {Schematic representation of a recurrence protocol. States form a successful purification round serve as inputs for the next purification round.}
\label{resourcesBEPP}
\end{center}
\end{figure}

Recurrence protocols operate in each purification step on a fixed number of copies of a mixed state. We will mainly consider recurrence protocols that operate on two identical copies here, but also treat briefly the cases where the copies are not identical (e.g. in pumping schemes) or more than two copies are involved. After local manipulation, one of the copies is measured, and depending on the outcome of the measurement the other copy is kept (we refer to this as a successful purification step) or discarded (see Fig. \ref{bipartitesetupFig}). In case of a successful purification step, the fidelity of the remaining pair is increased. The procedure is iterated, whereby states resulting from a successful purification round are used as input for the next purification round (see Fig. \ref{resourcesBEPP}). Typically, these protocols converge to a fixed point which is given by a maximally entangled state.

\subsubsection{Basic properties and notation}

We now turn to specific recurrence protocols that allow one to purify bipartite entangled states of two qubits. We will not describe these protocols as they were originally presented, but provide an equivalent description which will allow us a unified treatment of bipartite and multipartite entanglement purification protocols. In particular, we describe protocols that operate on states in a (locally) rotated basis and describe the corresponding states in terms of their stabilizing operators. To this aim, we start by fixing some notation. We consider two parties, $A$ and $B$, each holding several copies of noisy entangled states described by a density operator $\rho_{AB}$ acting on Hilbert space $\mathbb{C}^2 \otimes \mathbb{C}^2$. We denote by 
\begin{eqnarray}
|\Phi_{00}\rangle \equiv \frac{1}{\sqrt{2}}(|0\rangle_z|0\rangle_x\rangle + |1\rangle_z|1\rangle_x), 
\end{eqnarray} 
a maximally entangled state of two qubits, where $|0\rangle_z, |1\rangle_z$ [$|0\rangle_x, |1\rangle_x]$ are eigenstates of $\sigma_z$ [$\sigma_x$] with eigenvalue $(\pm 1)$ respectively. For example, $\sigma_x |1\rangle_x = -|1\rangle_x$, and $|0\rangle_x=1/\sqrt{2}(|0\rangle_z + |1\rangle_z)$. We also define
\begin{eqnarray}
|\Phi_{k_1k_2}\rangle \equiv \sigma_z^{k_1}\sigma_z^{k_2} |\Phi_{00}\rangle,
\end{eqnarray}
with $k_1,k_2 \in \{0,1\}$. The states $\{|\Phi_{k_1k_2}\rangle \}$ form a basis of orthogonal, maximally entangled states, the so called Bell basis.
We remark that the states $|\Phi_{k_1k_2}\rangle$ are joint eigenstates of correlation operators 
\begin{eqnarray}
K_1=\sigma_x^{(A)} \sigma_z^{(B)},  \hspace{1cm} K_2=\sigma_z^{(A)}\sigma_x^{(B)},
\end{eqnarray}
with eigenvalues $(-1)^{k_1}$ and $(-1)^{k_2}$ respectively. Whenever several copies of a mixed state are involved, we will refer to the different copies by numbers. For instance, $\rho_{A_1B_1}$ refers to the first copy of a state, while $\rho_{A_2B_2}$ refers to the second copy. In this case, party $A$ holds two qubits, $A_1$ and $A_2$.

We consider mixed states $\rho'_{AB}$ which we expand in the Bell basis,
\begin{eqnarray}
\label{depolli}
\rho'_{AB}= \sum_{k_1,k_2,j_1,j_2=0}^1 \lambda'_{k_1k_2j_1j_2} |\Phi_{k_1k_2}\rangle \langle \Phi_{j_1j_2}|.
\end{eqnarray}
This state can e.g. be created by sending the second qubit of a maximally entangled state through a noisy quantum channel ${\cal E}$, Eq. (\ref{channel}). One can always {\em depolarize} the state to a standard form by a suitable sequence of (random) local operations in such a way that the fidelity of the state, 
\be
F\equiv \langle\Phi_{00}|\rho_{AB}|\Phi_{00}\rangle
\ee  
is not altered. To be specific, by probabilistically applying one of the multi--local operations corresponding to $\{\one, K_1, K_2, K_1K_2\}$ one produces a density operator which is diagonal in the Bell basis,
\begin{eqnarray}
\rho_{AB}= \sum_{k_1,k_2=0}^1 \lambda_{k_1k_2} |\Phi_{k_1k_2}\rangle \langle \Phi_{k_1k_2}|,
\label{Bell}
\end{eqnarray}
and in which diagonal coefficients remain unchanged, $\lambda_{k_1k_2} \equiv \lambda'_{k_1k_2k_1k_2}$. This dephasing step can be understood as follows: Consider for instance the action of $K_1$ on basis states $|\Phi_{k_1k_2}\rangle$. For $k_1=0$, the state is left invariant while a global phase of $(-1)$ is picked up if $k_1=1$. It follows that off--diagonal elements of the form $|\Phi_{k_1k_2}\rangle\langle \Phi_{j_1j_2}|$ in (\ref{depolli}) are transformed to $(-1)^{k_1\oplus j_1} |\Phi_{k_1k_2}\rangle\langle \Phi_{j_1j_2}|$, i.e. acquire a phase if $k_1 \not= j_1$. Consequently, when applying the local operation $K_1$ with probability $p=1/2$ and leaving the state unchanged otherwise, the resulting density operator $\rho' = 1/2(K_1 \rho K_1^\dagger +\rho)$ has no off--diagonal elements where $k_1 \not= j_1$. In a similar way, all off--diagonal elements are cancelled by the (random) application of $\one,K_1,K_2,K_1K_2$. Note that all diagonal elements --in particular the fidelity of the state-- remain unchanged by this depolarization procedure. 
Using similar techniques, one can further depolarize the state by making all but one of the diagonal elements equal. The resulting states are called Werner states \cite{We89},
\begin{eqnarray}\label{Wernerstates}
\rho_W(x)= x|\Phi_{00}\rangle\langle\Phi_{00}| + (1-x)\frac{1}{4} \one_{AB},\label{rhoWerner}
\end{eqnarray}
where the fidelity $F=(3x+1)/4$ is unchanged. This can be accomplished by randomly applying local unitary operations that leave the state $|\Phi_{00}\rangle$ (up to a phase) invariant, which is the case for all operations of the form $U \otimes HU^*H$ with $H$ being the Hadamard gate \footnote{The Hadamard operation $H$ maps basis state of $z$ basis to basis states of $x$ basis and vice versa, i.e. $H|k\rangle_z = |k\rangle_x, H|k\rangle_x = |k\rangle_z$ with $k=0,1$.} and $^*$ denoting complex conjugation. The unitaries can be chosen uniformly (according to the Haar measure), or selected from a specific finite set of operations \cite{Be96a,Be96}. Notice, however, that the entanglement of the states may decrease by the depolarization procedure, even though the fidelity remains unchanged. What is important in our context is that any state with fidelity $F$ can always be brought to Werner form. It is thus sufficient to provide an entanglement purification method which works for Werner states, because such a method automatically allows one to purify {\em all} states which have the same fidelity, independent of their ``shape''. We consider such a purification procedure in the following.

\subsubsection{BBPSSW protocol}

In 1996, Bennett  et al. \cite{Be96a} introduced a purification protocol that allows one to create maximally entangled states with arbitrary accuracy starting from several copies of a mixed state $\rho$, provided that the fidelity $F$ with some maximally entangled state fulfills $F > 1/2$. The protocol consists of the following steps:
\begin{itemize}
\item[(i)] Depolarize $\rho$ to Werner form; 
\item[(ii)] apply bilateral local CNOT operations $U_{\rm CNOT}^{A_1 \to A_2} \otimes U_{\rm CNOT}^{B_2 \to B_1}$ \footnote{The CNOT operation is defined by $|i\rangle_A|j\rangle_B \rightarrow |i\rangle_A|i\oplus
j\rangle_B$, where $\oplus$ denotes addition modulo 2.}; 
\item[(iii)] measure qubit $A_2$ [$B_2$] locally in eigenbasis of $\sigma_z$ [$\sigma_x$] with corresponding results $(-1)^{\zeta_1}$ [$(-1)^{\xi_1}$] respectively, where $\zeta_1,\xi_1 \in\{0,1\}$. The effect of this local measurement on other particles is the same as the measurement of the observable $K_2^{(A_2B_2)}$; 
\item[(iv)] keep state of the pair $A_1B_1$ if $(\zeta_1+\xi_1){\rm mod}2=0$, i.e. measurement results coincide.
\end{itemize}
Given two copies of a state with fidelity $F$, it is straightforward to calculate the fidelity of the resulting state when applying (i-iv). The effect of (ii) on two Bell states is given by
\begin{eqnarray}
|\Phi_{k_1,k_2}\rangle_{A_1B_1} |\Phi_{j_1,j_2}\rangle_{A_2B_2} \rightarrow |\Phi_{k_1\oplus j_1,k_2}\rangle_{A_1B_1} |\Phi_{j_1,k_2\oplus j_2}\rangle_{A_2B_2}.
\label{actionBCNOT}
\end{eqnarray}
The effect of (iii) and (iv) is to select states in $A_2B_2$ which are eigenstates of $K_2^{A_2B_2}$ with eigenvalue (+1), while eigenstates with eigenvalue (-1) are discarded. That is, only initial states $|\Phi_{k_1,k_2}\rangle_{A_1B_1} |\Phi_{j_1,j_2}\rangle_{A_2B_2}$ with $k_2\oplus j_2 =0$ will pass the measurement procedure, which implies that, when considering mixed states, only these components will contribute to the final density operator. The final state turns out to be not of Werner form, however due to step (i) the state is brought back to Werner form when iterating the procedure. Hence the essential parameter is the fidelity $F'$ after a successful purification step. One finds
\begin{eqnarray}
F'=\frac{F^2+[(1-F)/3]^2}{F^2+2F(1-F)/3+5[(1-F)/3]^2},
\label{Fout}
\end{eqnarray}
which fulfills $F' >F$ for $F>1/2$. The success probability is given by the denominator of Eq. (\ref{Fout}), $p_{\rm suc}=F^2+2F(1-F)/3+5[(1-F)/3]^2$. Iteration of the procedure, which means to take two identical copies of states with fidelity $F'$, resulting from a previous, successful purification round, allows one to successively increase the fidelity. In fact, it is straightforward to see that the map Eq. (\ref{Fout}) has $F=1$ as an attractive fixed point. Hence states arbitrarily close to maximally entangled states can be produced. Even though the probability of success of the purification steps tends to one for $F \to 1$, the yield of the procedure goes to zero as one of the pairs is always measured and has to be discarded. Fixing however the desired target fidelity of resulting states to $F> 1-\epsilon_0$, a finite number of purification steps suffices and hence the yield will be finite. We remark that the obtainability of states with $F=1$ seems to be a question of only theoretical relevance, since imperfections in the apparatus used for the preparation of the state and in the purification procedure limit the reachable fidelity.

\subsubsection{DEJMPS protocol}

The DEJMPS protocol, introduced by Deutsch et al. in Ref. \cite{De96}, is conceptually similar to the BBPSSW protocol. It operates however not on Werner states, but on states diagonal in a Bell basis (see Eq. \ref{Bell}). The main advantage of this protocol is that it has higher efficiency. The protocol operates on two identical copies of a state and consists essentially of the same steps as the BBPSSW protocol. The only difference is that step (i) is replaced by a step (i'). 
\begin{itemize}
\item[(i')] perform local basis change 
\be
|0\rangle_z^{(A)} \to \frac{1}{\sqrt{2}} (|0\rangle_z^{(A)} -i|1\rangle_z^{(A)}),& \hspace{0.5cm} &|1\rangle_z^{(A)} \to \frac{1}{\sqrt{2}} (|1\rangle_z^{(A)} -i|0\rangle_z^{(A)}) \nonumber\\
|0\rangle_x^{(B)} \to \frac{1}{\sqrt{2}} (|0\rangle_x^{(B)} +i|1\rangle_x^{(B)}),& \hspace{0.5cm} &|1\rangle_x^{(B)} \to \frac{1}{\sqrt{2}} (|1\rangle_x^{(B)} +i|0\rangle_x^{(B)})\nonumber.
\ee 
\end{itemize}
The effect of step (i') is (up to some irrelevant phases) to flip the diagonal components of $|\Phi_{10}\rangle$ and $|\Phi_{11}\rangle$, i.e. $\lambda_{10} \leftrightarrow \lambda_{11}$. One may in addition add a depolarization of $\rho$ to Bell--diagonal form (see Eq. \ref{Bell}), however as shown in Ref. \cite{De96} the off--diagonal terms do not influence the protocol anyway. 
The total effect of the protocol (steps (i-iv)) can be described as a non-linear map for the diagonal components of $\rho$ to $\rho'$ (written in the Bell basis), i.e. a map from $\mathbb{R}^4 \to \mathbb{R}^4$. To be specific, the map reads \cite{De96}
\begin{eqnarray}
\lambda_{00}'=(\lambda_{00}^2+\lambda_{11}^2)/N, \hspace{1cm}&& \lambda_{01}'=(\lambda_{01}^2+\lambda_{10}^2)/N,\nonumber\\
\lambda_{10}'=2\lambda_{00}\lambda_{11}/N,\hspace{1cm} && \lambda_{11}'=2\lambda_{01}\lambda_{10}/N,
\label{Oxfordmap}
\end{eqnarray}
where $N=(\lambda_{00}+\lambda_{11})^2+(\lambda_{01}+\lambda_{10})^2$ is the probability of success of the protocol. Again, the protocol can be iterated, and the diagonal coefficients of the state in the Bell basis after $k$ successful purification steps can be calculated by $k$ iterations of the map Eq. (\ref{Oxfordmap}). One can show that the map has $\lambda_{00}=1, \lambda_{ij}=0$ for ${ij}\not=00$ as attracting fixed point, and in fact all states with $\lambda_{00}>1/2$ (i.e. $F>1/2$) can be purified \cite{Ma98}.

\subsubsection{Entanglement pumping}\label{nestedEPP}

While both the BBPSSW and DEJMPS protocol allow one to successfully produce entangled states with arbitrary high fidelity, the requirements on local resources are rather demanding. Since at every round two identical states resulting from previous successful purification rounds are required, the total number of pairs that have to be available initially increases exponentially with the number of steps. In real implementations these pairs have to be stored by some means. For many physical set--ups, however, the number of particles that can be stored is limited. 

\begin{figure}[ht]
\begin{center}
\includegraphics[width=0.85\textwidth]{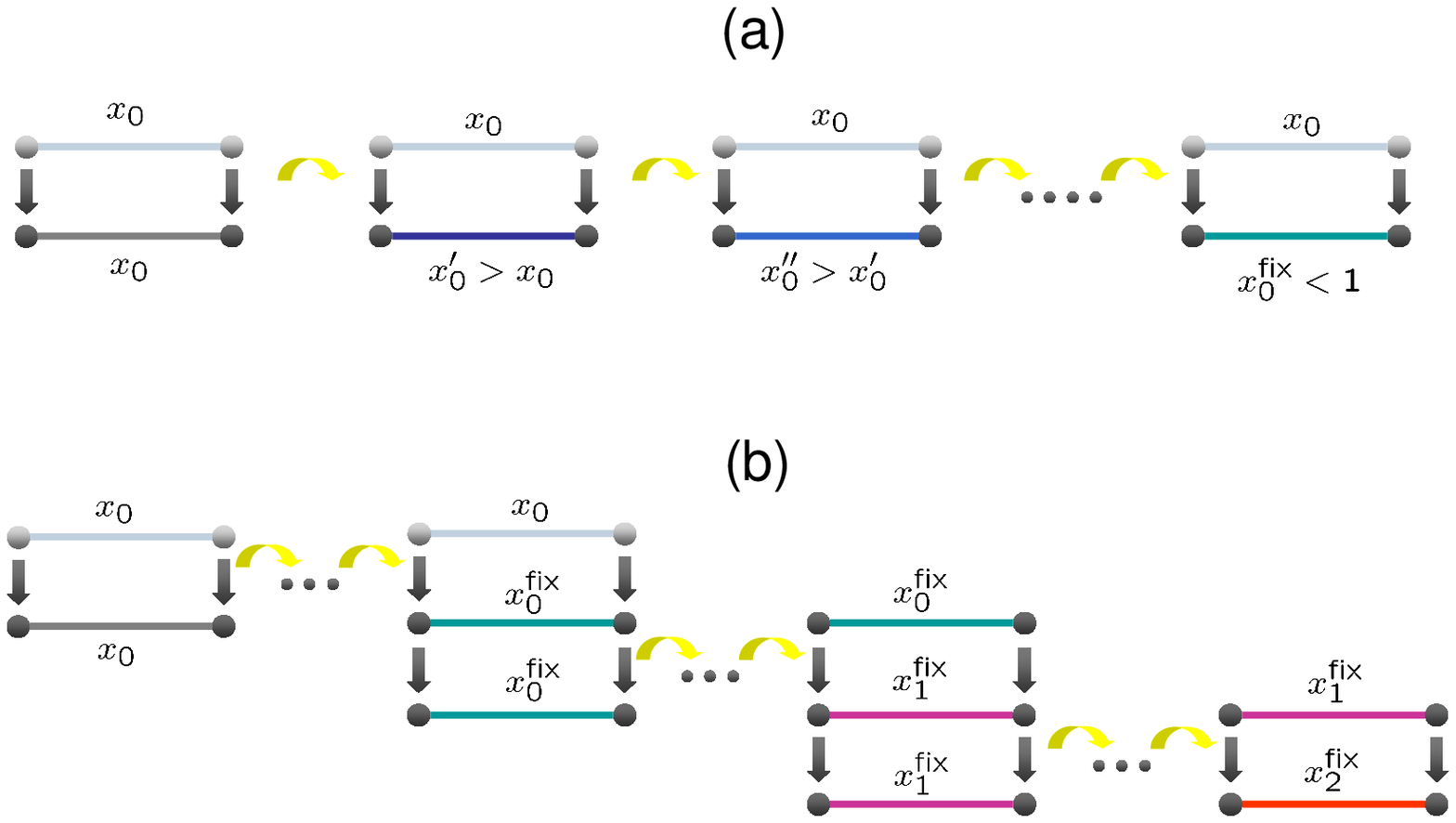}
\caption{(a) Schematic representation of entanglement pumping. An elementary pair (pair 1) with fidelity $x_0$ is repeatedly created and used to purify the second pair. The fidelity converges to some fixed point $x_0^{\rm fix} <1$. (b) Nested entanglement pumping. At nesting level 1, elementary pairs of fidelity $x_0$ are used to purify the second pair to fidelity $x_0^{\rm fix}$. At the next nesting level, pairs of fidelity $x_0^{\rm fix}$ produced in this way serve as elementary pairs and are used to purify one pair to fidelity $x_1^{\rm fix}$ via entanglement pumping. Pairs of fidelity $x_1^{\rm fix}$ are then used as elementary pairs at the next nesting level.}
\label{NestedPumping}
\end{center}
\end{figure}

The requirements in memory space can however be translated into temporal resources. The corresponding purification protocol is called (nested) entanglement pumping \cite{Br98,Du98,Du03QC} (see also Fig. \ref{NestedPumping}). The basic idea is to repeatedly produce elementary entangled pairs (resulting e.g. from the transmission of these maximally entangled state through noisy channels) and using a fresh elementary pair to purify a second pair. If a purification step is not successful, one has to start again from the beginning, using two elementary pairs. The actual sequence of local operations is either given by the BBPSSW or DEJMPS protocol, where the pair to be purified acts as pair 1 (source pair), while the fresh, elementary pair plays the role of pair 2 (target pair) that is measured. In case the purification step was successful, the fidelity of the first pair is increased by a certain amount. It is straightforward to determine the maps corresponding to Eqs. (\ref{Fout},\ref{Oxfordmap}) for non--identical input states. One finds
\begin{eqnarray} F'=\frac{F_1F_2+\left(\frac{1-F_1}{3}\right)\left(\frac{1-F_2}{3}\right)}{F_1F_2+F_1\left(\frac{1-F_2}{3}\right)+\left(\frac{1-F_1}{3}\right)F_2+5\left(\frac{1-F_1}{3}\right)\left(\frac{1-F_2}{3}\right)}. \label{Fout2}
\end{eqnarray}
in the case of two Werner states with fidelity $F_1,F_2$. In this map, $F_2$ is to be considered as constant since the second pair is always an elementary one. For two Bell diagonal states with coefficients $\lambda_{ik}$ and $\mu_{ik}$ we obtain 
\begin{eqnarray}
\lambda_{00}'=(\lambda_{00}\mu_{00}+\lambda_{11}\mu_{11})/N, && \lambda_{01}'=(\lambda_{01}\mu_{01}+\lambda_{10}\mu_{10})/N,\nonumber\\
\lambda_{10}'=(\lambda_{00}\mu_{11}+\lambda_{11}\mu_{00})/N, && \lambda_{11}'=(\lambda_{01}\mu_{10}+\lambda_{10}\mu_{01})/N,
\label{Oxfordmap2}
\end{eqnarray}
Again, the second pair is always an elementary one, and hence $\mu_{ik}$ is fixed. While iteration of the corresponding maps allows in both cases to improve the fidelity, {\em no} maximally entangled states can be generated, in general. That is, the fixed point of the maps described by Eqs. (\ref{Fout2},\ref{Oxfordmap2}) depends on the coefficients $\mu_{ik}$ and specifically on the fidelity of the elementary pair \cite{Du98}.

As elementary pairs can be generated on demand, they do not need to be stored. Hence in $A$ and $B$ only two qubits need to be stored (corresponding to the pair to be purified and the elementary pair, respectively). The reduction in spatial resources leads however to an increase of temporal resources. In protocols BBPSSW and DEJMPS, the purification of different pairs corresponding to a single purification step can be implemented in parallel (i.e. the temporal resources are given by the number of steps). That is, one actually considers a distillation procedure where out of many low--fidelity entangled pairs a few with higher fidelity are generated. The probabilistic character of entanglement purification manifests itself in the fact that many identical pairs need to be simultaneously available. In entanglement pumping, in contrast, the probabilistic character of purification leads to increased number of required repetitions, as in case of an unsuccessful purification step the procedure has to be started from beginning and pairs are {\em sequentially} generated.

One can improve the entanglement pumping scheme in such a way that the number of qubits that have to be locally stored remain small ($\approx 4$ for practical purposes), while it is possible to generate maximally entangled states rather than only enhancing the fidelity by a finite amount. The corresponding scheme is called {\em nested entanglement pumping} \cite{Du03QC} and works as follows: At nesting level 1, elementary pairs created between $A_1$ and $B_1$ are used to purify a pair shared between $A_2$ and $B_2$ via entanglement pumping. The fidelity of elementary pairs at level 1 is given by $F_1$. It turns out that after a few purification steps, the fidelity $F_2$ of the pair $A_2$ and $B_2$, is already close to the reachable fixed point. The resulting pair with improved fidelity $F_2$ now serves as elementary pair at nesting level 2. That is, an elementary pair at nesting level 2 shared between $A_3$ and $B_3$ is purified by means of entanglement pumping, where always elementary pairs of nesting level 2 with fidelity $F_2$ shared between $A_2$ and $B_2$ are used. The fidelity of the resulting pair $A_3$ and $B_3$ after a few purification steps is given by $F_3$ with $F_3>F_2>F_1$. We remark that an unsuccessful purification step at a higher nesting level requires to restart the procedure at the lowest nesting level 1. Still, the required temporal resources increase only polynomially. The overall procedure can be viewed as a stochastic process, or equivalently as a one--sided bounded random walk.
With each nesting level, one additional particle has to be stored at each location. However, it turns out that for practical purposes (say required accuracy of $\epsilon_0=10^{-7}$) a few nesting levels ($\approx 3$) suffice to generate states with fidelity $F > 1-\epsilon_0$ \cite{Du03QC}. Hence the storage requirements remain very moderate, while the required temporal resources increase only polynomially.

\subsection{$N\to M$ protocols}

The protocols discussed in the previous Section operate on two copies of a given mixed state, and produce one copy as output if they are successful. More general protocols are conceivable that operate on $N$ input copies of the state and produce $M$ copies as output. We will refer to such protocols as $N \to M$ protocols, and discuss them in this subsection. A protocol of this kind of particular importance is the so--called hashing protocol, which operates in the limit $N,M \to \infty$. The general idea behind $N \to M$ protocols is very similar as in the case of standard recurrence protocols operating on two copies: To obtain information about a sub--ensemble --in this case consisting of $M$ copies of the state--, for which the remaining $N-M$ copies are measured after applying suitable local operations. 

\subsubsection{$N \to M$ protocols for finite $N$}

The $2 \to 1$ recurrence protocols discussed in the previous section can be considered as two--step procedures. In a first step, the two copies of the input state are manipulated by local operations that entangle the two pairs. The effect of these local operations on Bell diagonal states is a certain {\em permutation} of the basis elements. In a second step, the second pair is measured, and depending on the outcome of the measurement the first pair is either kept or discarded. 
General $N \to M$ protocols operate in a very similar fashion. In fact, in Ref. \cite{De01} all possible permutations achievable by local operations have been constructed for qubit systems, and accordingly a large number of possible $N \to M$ entanglement purification protocols were constructed and analyzed. It was found that in certain regimes such $N \to M$ protocols operate more efficiently (i.e have a higher yield) than standard $2 \to 1$ protocols \cite{De01,Ma01}. Typically, for small initial fidelities the ratio of final pairs $M$ to initial pairs $N$ may be small, $M/N \ll 1$, while one expects that $M/N \approx 1$ for large fidelities as only a small amount of information about the remaining ensemble needs to be revealed.
Generalizations of this concept to the purification of entangled $d$--level systems are possible \cite{Bo05} (see also e.g. Refs. \cite{Ho99,Al01,Ch05} for entanglement purification protocols of $d$--level systems).

We would also like to remark that a general connection between error correcting (stabilizer) codes and $N \to M$ purification protocols exists \cite{AsPhD}, which we describe later in more detail. In fact, for each code one can construct a corresponding $N \to M$ entanglement purification protocol.

\subsubsection{Hashing and breeding protocols}

Hashing protocols can be considered as special instances of $N \to M$ protocols that operate in the limit $N \to \infty$. Hashing was introduced in Ref. \cite{Be96}. The basic idea is similar as in $N \to M$ recurrence protocols. Here, random subsets of size $n$ out of the total $N$ copies of the state are chosen, and bilateral local CNOT operations with each of the $n$ pairs as source, and one selected pair as target, are performed (or vice versa, i.e. the selected pair as source). The selected pair is finally measured, revealing one bit of information about the remaining ensemble. Measurements of this kind are repeated $m$ times. One can in fact show that the information gain per measurement is close to one bit. 

Hashing is conceptually closely related to breeding, which might be slightly easier to understand. In the case of breeding, the parties are assumed to possess, in addition to the $N$ copies of the state, $m$ pre-purified, maximally entangled Bell pairs which are used to gain information about the remaining ensemble. 
In the asymptotic limit of large $N$ the density matrix $\rho^{\otimes N}$ is approximated to an arbitrary high accuracy by its ``likely subspace approximation'', i.e. the density matrix $\Gamma$ obtained by projecting $\rho^{\otimes N}$ into a subspace $P$ (the likely subspace), where the dimension of $P$ is $2^{(S(\rho)+\delta)N}$. In the case of Werner states $\rho_W(F)$ (see Eq. \ref{rhoWerner}, $F=(3x+1)/4$), this likely subspace contains essentially states of the form $\otimes_{i,j} |\Phi_{ij}\rangle^{\otimes m_{ij}}\langle \Phi_{ij}\rangle^{\otimes m_{ij}}|$ and permutations thereof, where $m_{00}= F N, m_{01}=m_{10}=m_{11} =(1-F)/3 N$ 
\footnote{To be precise, one has to consider in addition also states with $m_{00}=(F N \pm O(\sqrt N))$ and similarly for $m_{ij}$.}. That is, the density matrix $\rho^{\otimes N}$ can be interpreted as an equal mixture of all these possible configurations, where the number of Bell states $|\Phi_{ij}\rangle$ is essentially fixed to $m_{ij}$, while the order (or position) of the states is unknown. The number of possible configurations of states of this form is --for large $N$-- approximately given by $2^{N S(F)}$, where $S(F)= -F\log_2F-(1-F) \log_2(\frac{1-F}{3})$. The task thus reduces to reveal which of these possible configurations one is dealing with. Clearly, this requires $N S(F)$ bits of information. Since one can gain at most one bit of information about the ensemble with help of each maximally entangled pair, one needs $m=N S(F)$ additional maximally entangled pairs to perform this task. Having obtained the required information, one possesses a pure state consisting of $N$ Bell states (in different bases), i.e. some (known) permutation of the state $\otimes_{i,j} |\Phi_{ij}\rangle^{\otimes m_{ij}}\langle \Phi_{ij}\rangle^{\otimes m_{ij}}|$. Since $m=S(F)N$ maximally entangled pairs have been consumed during the process, the total yield of the breeding protocol is given by $D=1-S(F)$. Note that $S(F)= S(\rho_W)$, where $S(\rho_W) = -{\rm tr}(\rho_W \log_2 \rho_W)$ is the von-Neumann entropy of $\rho_W$. It follows that for Werner states, breeding only works if the initial fidelity is sufficiently high, $F \gtrsim 0.81$.      

A similar kind of reasoning can be applied to hashing, where no pre--purified pairs are required. The analysis is slightly more involved since one has to take a kind of back action (influence of the remaining pairs because the measured pair was not in a pure state) into account. The yield of the hashing procedure is, however, exactly the same as for breeding. For Bell--diagonal states, one obtains that the yield of hashing protocols is given by $D(\rho) = 1-S(\rho)$. 

The yield of hashing and breeding protocols can be further improved if two--way classical communication is allowed, see e.g. Ref. \cite{Ve04,Ho06}. The underlying principle for this improvement is discussed in Ref. \cite{Ho06}. In addition, one can generalize hashing and breeding to $d$ dimensional systems for prime $d$ \cite{Vo04}. 
The optimal entanglement distillation protocol for two--way classical communication is in general unknown (see however \cite{Ho04,Gh04}). Only for specific two qubit states, for instance incoherent mixtures of two Bell states, the known upper bounds on the yield coincide with the achievable rate for known protocols, in this case the hashing protocol. When assuming only one--way classical communication, the problem becomes tractable. In fact, the optimal distillation protocol for one--way classical communication was obtained in Ref. \cite{De04}.


\section{Quantum error correction \& entanglement purification}\label{relation}

The two possible solutions to the problem of quantum communication over noisy channels, quantum error correction and entanglement purification plus teleportation, have already been discussed in Sec. \ref{ECC}. Here we show that the approaches are in fact equivalent when considering one--way classical communication. In addition, we discuss a systematic way how to construct entanglement purification protocols from quantum error correction codes. We treat not only the (straightforward) case of one--way protocols, but also protocols that make use of two--way classical communication. 

\subsection{Equivalence between QECC and one-way EPP}

The close relation between quantum error correction codes and schemes based on one--way classical communication has been proven in an early paper \cite{Be96}. Consider a quantum channel ${\cal E}$ and the associated bipartite mixed state 
\be\label{EfromcalE}
\hat E = \eins \otimes {\cal E} |\Phi\rangle\langle \Phi|
\ee 
that is obtained by sending part of a maximally entangled state $|\Phi\rangle$ through the channel ${\cal E}$. Consider also the channel $\hat{\cal E}$ which is generated when using a state $E$ for teleportation. Notice that for Pauli--diagonal channels ${\cal E_P}$ (see Eq. \ref{Paulidiag}) we have $\hat{\cal E_P} = {\cal E_P}$ and $\hat E_P =E_P$. 

In Ref. \cite{Be96} two inequalities are shown which establish this relation: 
\begin{itemize}
\item[(i)] $D_E^{\rightarrow} \geq Q^\rightarrow_{\hat{\cal E}}$
\item[(ii)] $D_{\hat E}^{\rightarrow} \leq Q^\rightarrow_{{\cal E}}$
\end{itemize}
The two inequalities are proven by establishing explicit protocols. Regarding (i), one considers the QECC which leads to channel capacity $Q^\rightarrow_{\hat{\cal E}}$. The second particles of $m$ locally prepared maximally entangled pairs are encoded using the QECC and teleported to $B$ using several copies of the mixed state $E$. At $B$, the decoding operation is applied and errors are corrected. This leads to $m$ maximally entangled states, and we have in fact described for any $E$ a one--way entanglement purification protocol which reaches equality in (i), and hence (i) is fulfilled. 

Regarding (ii), one considers the entanglement purification protocol ${\cal P}$ that leads to $D_{\hat E}^{\rightarrow}$. One creates several copies of the bipartite mixed state ${\hat E}$ by sending the second particle of a maximally entangled states through the channel, and uses ${\cal P}$ to generate maximally entangled pairs with rate $D_{\hat E}^{\rightarrow}$. These pairs are then used for (perfect) teleportation, and we have constructed in this way a coding scheme which reaches equality in (ii), and hence (ii) is fulfilled. 

Notice that in the case of Pauli--diagonal channels (and also for some other channels), (i) and (ii) show the {\em equivalence} between one--way entanglement purification and quantum error correction in the sense that one--way distillable entanglement and channel capacity are the same,
\be
D_E^\rightarrow = Q^\rightarrow_{{\cal E}}.
\ee

\subsection{One-- and two--way entanglement purification protocols from CSS codes}\label{EPPfromQECC}

For Pauli--diagonal noise channels (Eq. \ref{Paulidiag})), also a direct way how to derive a QECC from certain one--way entanglement purification protocols has been established in Ref. \cite{Be96}. One can turn the construction around, which leads to a constructive way of obtaining one--way entanglement purification protocols from a certain class of QECC, the CSS codes \cite{Ca96,St96}. This approach can also be generalized, as shown by Aschauer in his PhD thesis \cite{AsPhD}, in the sense that not only entanglement purification protocols using one--way classical communication, but also protocols making full usage of two--way classical communication can be constructed from CSS codes (see also Refs. \cite{Ma02,Am03} for alternative approaches). In particular, for each CSS code that uses $n$ physical qubits to encode $k$ qubits, one can construct an entanglement purification protocol that operates on $n$ initial copies of two-qubit states and produces $k$ purified pairs as output.

There are two possible operational modes for such entanglement purification protocols
\begin{itemize}
\item[(i)] error correction mode
\item[(ii)] error detection mode
\end{itemize}
In case of (i) only one-way classical communication is used, and the $k$ output pairs are kept deterministically. Measurements on the remaining $n-k$ copies are used to gain information on the ensemble, and the measurement outcomes determine the error correction operation to be applied to the remaining copies. As shown in Ref. \cite{AsPhD}, entanglement purification protocols can also run in an alternative mode (ii) where one makes use of two--way classical communication. The information gathered in the measurement of the $(n-k)$ pairs is used to decide whether the remaining pairs should be kept or discarded. The ones that are kept have a higher fidelity than before. This operational mode is the standard mode for recurrence protocols as discussed above, and in fact turns out to provide a larger purification range and favorable error thresholds.  

\begin{figure}
\begin{center}
\includegraphics[width=0.85\columnwidth]{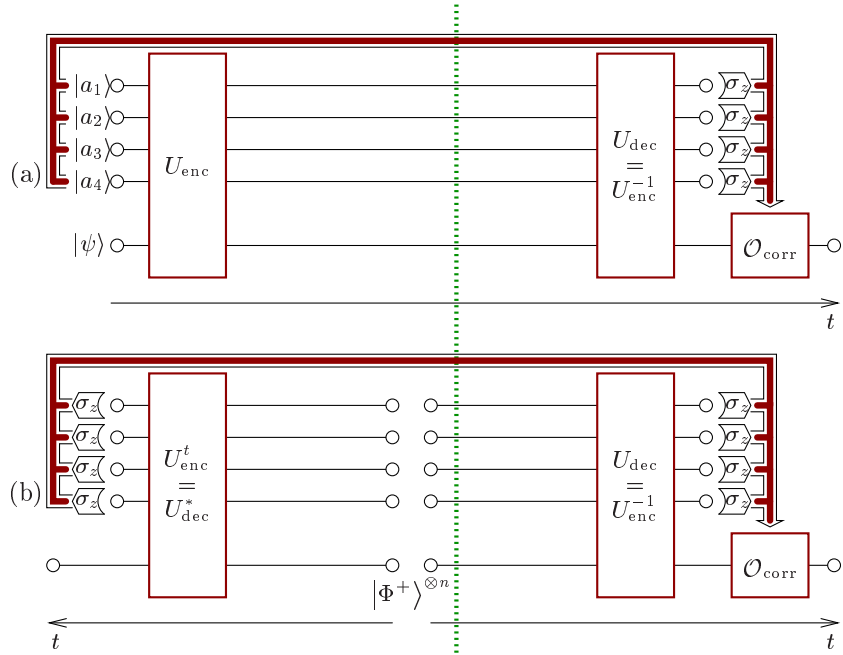}
\caption {Schematic representation of how to construct an entanglement purification protocol from an error correction CSS code. Figure taken from \cite{AsPhD}.}
\label{AsFig}
\end{center}
\end{figure}

We now illustrate this construction (for details we refer the reader to Ref. \cite{AsPhD}; see also Fig. \ref{AsFig}). To this aim, consider a Pauli--diagonal channel ${\cal E_P}$ which we will denote simply by ${\cal E}$ here. Consider also a mixed state $E$ which is obtained by sending one particle of a maximally entangled state $|\Phi\rangle$ through the channel, Eq. (\ref{EfromcalE}). For any CSS code described by coding operations ${\cal C}$ and decoding operations ${\cal D}$, it is shown how to obtain an entanglement purification protocol. Notice that ${\cal C},{\cal D}$ are Clifford networks (i.e. composed only of Clifford operations), where ${\cal C}$ consists of an encoding unitary operation $U_C$, while the decoding operation is given by a unitary operation $U_D = U_C^\dagger$ followed by single qubit measurements in the eigenbasis of $\sigma_z$ on the last $n-1$ qubits. 
We make use of the identity
\be\label{phipluseq}
U_A\otimes \one_B \ket{\Phi^+}_{AB}=\one_A\otimes U_B^T \ket{\Phi^+}_{AB},
\ee
which holds not only for all unitary operations $U$, but in fact for any linear operator. 

We consider first a QECC where one qubit is encoded into $n$ physical qubits. The transmission of (unknown) quantum information from $A$ to $B$ using such a QECC takes place by encoding the unknown state at $A$ into $n$ qubits, sending it through the noisy channel ${\cal E}^{\otimes n}$ and decoding it at $B$, where the decoding operation also includes error correction. Let $\rho_{\rm in}$ the state to be sent and denote by $P_{\bm 0}= |0\rangle^{\otimes n-1}\langle 0|$ the state of $n-1$ auxiliary qubits. The final state of this procedure is described as
\be\label{finalstate}
\rho_{\rm out} = {\cal D}\circ {\cal E}^{\otimes n} \circ {\cal C} (\rho_{\rm in}\otimes P_{\bm 0}). 
\ee 

Consider now an entanglement--based version of the protocol, which involves the following steps:
\begin{itemize}
\item[(i)] distribute $n$ copies of a maximally entangled state $|\phi^+\rangle$ through the noisy quantum channel ${\cal E}^{\otimes n}$, 
\item[(ii)] apply the coding operation ${\cal C}^T$ in $A$, and the decoding operation ${\cal D}$ in $B$. The coding operation ${\cal C}^T$ is defined as the application of the unitary operation $U_C^T$ followed by single qubit $\sigma_z$ measurements on the last $n-1$ qubits, and similarly for ${\cal D}$. 
\item[(iii)] use the resulting single entangled pair to teleport the unknown state $\rho_{\rm in}$ from $A$ to $B$, where the final correction operations not only depends on the result of the Bell measurement in the teleportation protocol, but also on the measurement outcomes on the qubits in $A$ and $B$ of the last $n-1$ entangled pairs.
\end{itemize} 

Consider first the case where ${\cal E} = \hat{\eins}$, i.e. a noiseless quantum channel. One can use the identity Eq. (\ref{phipluseq}) to show that one ends up with a maximally entangled pair that can be used for perfect teleportation. In a similar way, the equivalence of the QECC protocol and the entanglement based protocol can be established, where in the latter case the proper (Pauli) correction operations need to be applied. To establish this equivalence, it is essential that the noisy channels are Pauli--diagonal and that the coding and decoding operation are of Clifford type, because in this case the order of all the mentioned operations can be exchanged (up to additional, correctable local Pauli operations) and one can make use of Eq. (\ref{phipluseq}). In fact, one finds for such coding and decoding operations that the output state of the entanglement--based version of the protocol is again given by Eq. (\ref{finalstate}). 

Steps (i) and (ii) of the above construction (together with a suitable correction operation) can be interpreted as an {\em entanglement purification protocol}. This is due to the fact that an entangled state with improved fidelity can be created in this way whenever the QECC allows one to reduce the influence of channel noise. Using standard error correcting CSS codes, one obtains one--way entanglement purification protocols, as the measurement results are only used to determine the appropriate correction (Pauli) operation. However, here one has also the freedom to {\em discard} the resulting pair for certain measurement outcomes, which corresponds to the usage of an {\em error detection code} \cite{AsPhD}. Information exchange between $A$ and $B$ of the measurement outcomes is required, and hence two--way classical communication is involved. Notice that in entanglement purification one is allowed to discard certain pairs. This is in contrast to the direct transmission of quantum information, where discarding the state would yield to a loss of quantum information.  One can simply repeat the purification protocol until one succeeds in producing a purified output pair. Only then the (unknown) quantum information is teleported from $A$ to $B$. On the other hand, when sending encoded quantum information directly through the noisy channel, usage of error detection codes leads to unrecoverable {\em loss} of quantum information. Hence two--way entanglement purification protocols constructed in this way are superior to quantum communication schemes based on QECC.

Notice that the same construction can also be applied for CSS codes where $k$ qubits are encoded into $n$ physical qubits, and one obtains entanglement purification protocols which produces from $n$ noisy entangled pairs $k$ purified pairs \cite{AsPhD}.


\section{Entanglement purification with imperfect apparatus}\label{realistic}

In this section, we investigate the performance of entanglement purification protocols under non--idealized conditions, i.e. for noisy local control operations. The main effect of noise is that no longer maximally entangled states can be produced, and the achievable fidelity is limited to values smaller than unity. Similarly, the required initial fidelity in the case of noisy local control operations is larger. While recurrence protocols remain applicable, hashing and breeding protocols become impractical.

\subsection{Distillable entanglement and yield}

Using the standard definition of distillability and yield is clearly inappropriate in the case of imperfect local operations. In particular, no maximally entangled pure states can be created in this case. This implies that {\em no} state will be distillable, and that the yield is zero. We therefore have to modify the definition of distillability and yield to account for these facts.

Rather than demanding that maximally entangled pure states can be created (fidelity $F=1$), we will consider the creation of states with certain {\em target fidelity}. Distillability refers in this case to the possibility of approximating a given target state $|\psi\rangle$ with fidelity $F \geq F_c$. Clearly, such a definition of distillability depends on both the required target state $|\psi\rangle$ and the desired fidelity $F_c$. To be more precise, we say that a given mixed state $\rho$ is distillable with respect to a target state $|\psi\rangle$ and fidelity $F_c$ if one can generate from possibly many copies of $\rho$ by means of local operations and classical communication a state $\sigma$ such that the fidelity of $\sigma$ with respect to $|\psi\rangle$ is larger or equal than $F_c$, $F=\langle \psi|\sigma|\psi\rangle \geq F_c$. 

We consider the yield of purification procedures corresponding to this notion of distillability, $D_{\rho,F_c}$. In this case, however one needs to specify the exact structure of target states. In particular, when considering general distillation procedures (e.g. $N \to M$ protocols), one obtains as output a mixed state $\Gamma$ of a large number of particles. Here we will demand that the output state $\Gamma$ is a {\em tensor product} of states $\sigma_k$, $\Gamma = \sigma_1 \otimes \ldots \otimes \sigma_M$, where each of the $\sigma_k$ fulfills $\langle \psi|\sigma_k|\psi\rangle \geq F_c$. That is, we require that after the purification procedure one possesses independent copies of the state with desired fidelity. One may also use the weaker criterion that all reduced density operators $\tilde\sigma_k$ (corresponding to different output ``copies'' of the output state) have fidelity $F \geq F_c$, where $\tilde \sigma_k$ are obtained from $\Gamma$ by tracing out all particles but the ones corresponding to state $k$. In this case, however, it is not clear whether the different output states can be independently used for all applications. While their fidelities certainly fulfill $F \geq F_c$, there might be classical correlations among the output states that are limiting their applicability, e.g. for security applications such as key distribution.

In this context it would be interesting to see whether the definition of yield with respect to fidelities of reduced density operators is equivalent to the one we use here. To this aim, one would need to show that one can produce from an ensemble of states where all reduced density operators have a sufficiently high fidelity an ensemble which consists of a tensor product of copies, where the size of the ensembles might be diminished by a sub--linear amount, or the fidelity be reduced by some (arbitrarily small) $\delta_F$. Such a ``purification of classical correlations''  has, however, not been reported so far.

\subsection{Error model}\label{Errormodel}

To analyze the influence of noisy local operations, we will consider a simple error model where only two--qubit operations are noisy, and the noise is of a simple form. More general error models, including correlated noise and also errors in measurements, have been analyzed, leading to a similar  qualitative behavior of entanglement purification protocols \cite{Du98,Du03QC,Gi99}.

We model a noisy two--qubit operation $U$ by first applying local noise to each of the qubits, followed by the perfect unitary operation $U$,
\begin{eqnarray}\label{noisygate}
{\cal E}_{kl} \rho= U_{kl} [{\cal M}_k{\cal M}_l \rho ] U_{kl}^\dagger.
\end{eqnarray}
We will mainly assume local completely positive maps ${\cal M}_k$, ${\cal M}_l$ corresponding to white noise (depolarizing channels),
\begin{eqnarray}
{\cal M}_k \rho = p\rho + (1-p)\frac{1}{4} \sum_{j=0}^3 \sigma_j^{(k)} \rho \sigma_j^{(k)}\label{localmap}.
\end{eqnarray}
In some cases, we will consider even more restricted noise models, namely local dephasing channels (or phase flip channels),
\be
{\cal M}_k^P \rho = p\rho + (1-p)\frac{1}{2} (\rho+ \sigma_3^{(k)} \rho \sigma_3^{(k)}),
\ee
and local bit--flip channels, 
\be
{\cal M}_k^B \rho = p\rho + (1-p)\frac{1}{2} (\rho+ \sigma_1^{(k)} \rho \sigma_1^{(k)}).
\ee

\subsection{Bipartite recurrence protocols}

\begin{figure}[ht]
\begin{center}
\includegraphics[width=0.65\textwidth]{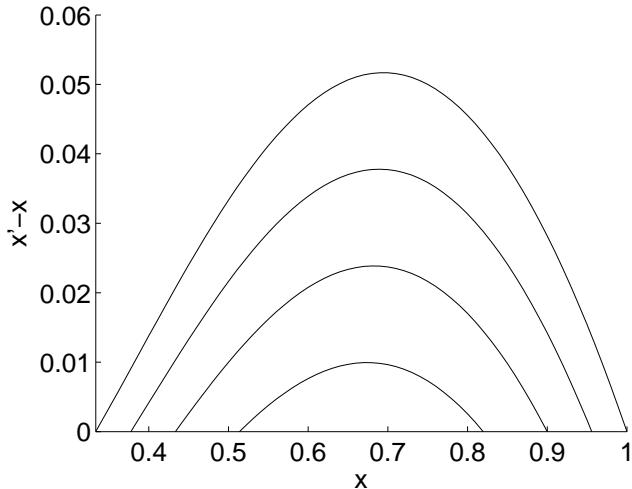}
\caption[]{\label{purificationcurve} Purification curve for BBPSSW protocol. Gain in output fidelity $x'-x$, plotted against input fidelity $x$. Curves from top to bottom correspond to error parameters $p=1, 0.99, 0.98, 0.97$ respectively.}
\end{center}
\end{figure}

We start by analyzing the BBPSSW protocol, where we consider the error model specified in Sec. \ref{Errormodel} with local white noise, Eq. (\ref{localmap}). Given two copies of a Werner state Eq. (\ref{rhoWerner}), the influence of noisy local control operations ---in this case noisy CNOT operations--- can be readily obtained. The action of noisy bilateral CNOT operations is the same as applying noiseless bilateral CNOT operations to two copies of Werner states with reduced fidelity. In particular, one finds that the parameter $x$ is mapped to $xp^2$ due to the local depolarizing noise. That is, one applies the original protocol to two copies of Werner states $\rho_W(xp^2)$. Rewriting Eq. (\ref{Fout}), i.e. the fidelity of output state as function of input state, in terms of parameter $x=(4F-1)/3$, one obtains $x'=(4x^2+2x)/(3x^2+3)$. Taking into account the effect of noisy local operations, i.e. the reduction of $x$, we obtain that the output state after one purification step is again a Werner state $\rho_W(x')$ with
\begin{eqnarray}
x'=\frac{4x^2p^4+2xp^2}{3x^2p^4+3}.\label{imperfectEPP}
\end{eqnarray}
The purification curve (fidelity of output state plotted against fidelity of input state) corresponding to the noiseless protocol is shifted down (see Fig. \ref{purificationcurve}).

It is now straightforward to determine the purification range of the protocol, i.e. maximal reachable fidelity as well as the minimal required fidelity such that entanglement purification can be successfully applied. These quantities are given by the fixed points of the map Eq. (\ref{imperfectEPP}). One finds
\begin{eqnarray}
x_\pm= \frac{2}{3} \pm \frac{1}{3} \sqrt{4 +6 p^{-2} - 9 p^{-4}},
\end{eqnarray}
where the maximum reachable fidelity $F_{\rm max}=(3x_+ +1)/4$ and the minimum required fidelity $F_{\rm min}=(3x_- +1)/4$. The threshold value for $p$ such that a finite purification interval remains (i.e. $x_+ > x_-$) is given by $p_{\min}=0.9628$. This implies that errors of the order of four percent are tolerable.

One can perform a similar analysis for the DEJMPS and (nested) entanglement pumping protocol. There, the fixed points of the corresponding non--linear maps are more difficult to obtain analytically. One can, however, perform the analysis numerically and obtains the following results \cite{Du98,Du03QC}  
\begin{itemize}
\item[(i)] the maximum reachable fidelity $F_{\rm max}$ for the DEJMPS protocol is significantly higher than for the BBPSSW protocol; 
\item[(ii)] the minimal required fidelity $F_{\rm min}$ for the DEJMPS is significantly smaller than for the BBPSSW; 
\item[(iii)] the threshold for noisy operations, described by $p_{\rm min}$ is smaller for the DEJMPS protocol;
\item[(iv)] the reachable fidelity, minimum required fidelity and threshold for noisy operations seem to be the same for nested entanglement pumping and for the original DEJMPS protocol \cite{Du03QC}.
\end{itemize}
When assuming errors in local operations corresponding to correlated white noise and errors in measurements of same order of magnitude \cite{Du98}, one finds tolerable errors of about three percent for the BBPSSW protocol, and five percent in case of the DEJMPS protocol.

\subsection{$N \to M$ protocols}

In a similar way, one can analyze the influence of noise for general $N \to M$ protocols. In Ref. \cite{AsPhD}, a number of protocols corresponding to different CSS codes (via the construction presented in Sec. \ref{EPPfromQECC}) have been analyzed with respect to their purification range, maximal reachable fidelities and error thresholds. Both operational modes, error correction mode and error detection mode have been considered. As can be expected due to the equivalence to QECC, the correction modes turn out to have a smaller purification range and much stringent error thresholds. When compared to standard recurrence protocols that operate on two copies, some of these $N \to M$ purification protocols turn out to have an improved yield in certain regimes.

\subsection{Hashing protocols}

For perfect local operations recurrence protocols have zero yield, while hashing protocols, operating simultaneously on an asymptotically large number of copies, have a non--zero yield. For imperfect local operations, the situation changes drastically. When requiring output states to have only a certain $F \geq F_c$, one finds that recurrence protocols may have a non--zero yield as long as $F_c \leq F_{\rm max}$, i.e. as long as the required fidelity is smaller than the fidelity reachable by the protocol. At the same time, the hashing protocol fails completely in the case of imperfect local operations. The reason for this is that one operates on an infinite number of states $m \to \infty$ to reveal one bit of information. That is, one performs $m$ bilateral CNOT operations with a given copy always serving as target state. As each of the CNOT operations is noisy, noise is accumulated in the target state. Assuming that the target state was initially in a maximally entangled pure state, the target state ends up in a Werner state $\rho_W(p^{2m})$. Clearly, if the amount of noise is too big, no information about the remaining ensemble can be extracted. This is the case for sufficiently large $m$, in particular for $m \to \infty$, even if $p$ is close to 1. In other words, the information loss due to imperfect local operations exceeds the possible information gain per measurement (maximum one bit). This implies that hashing in its original form can not be applied in the case of imperfect local operations.


\section{Applications I}\label{applications1}

\subsection{Quantum communication and cryptography}

As discussed in Sec. \ref{ECC}, entanglement purification together with teleportation offers a way to achieve perfect transmission of unknown quantum information over noisy channels. This approach can also be used for quantum key distribution in the context of quantum cryptography. 

In the case where not only the channels but also the local operations are imperfect, entanglement purification can still be applied. As we have seen in the previous section, one can increase the fidelity of entangled states --and hence the quality of the channel when using the purified entangled states for teleportation. More importantly, the entanglement produced by entanglement purification, although not perfect, is {\em private} \cite{As02}. That is, although no maximally entangled states can be produced, any eavesdropper will be factored out asymptotically. Hence a secret key can be established between two parties, even in the presence of noisy channels and imperfect apparatus \cite{As02}. This provides an alternative way of proving unconditional security of quantum key distribution, and is an important application of entanglement purification for quantum cryptography. We also mention that a direct and rigorous proof of the security of quantum cryptography that makes use of entanglement purification has been put forward (see e.g. Refs. \cite{Sh00,Go03,Ra06}).

\subsection{The quantum repeater}

In the context of quantum communication, a central problem is to go to large distances. When using photons, the absorption probability scales exponentially with the distance, and so does the influence of errors e.g. due to dephasing. This is a general, unavoidable feature of classical and quantum communication channels, and hence has to be dealt with in some way. The standard classical approach of amplifying the signal at intermediate repeater stations cannot be directly adopted in the case of quantum communication. This is due to the fact that general quantum signals, i.e. unknown quantum states, cannot be cloned or amplified.  

One possible solution is given by entanglement--based communication schemes, where entangled pairs are distributed over noisy quantum channels and then used to transmit arbitrary quantum information via teleportation. A nested sequence of connection (i.e. entanglement swapping) and entanglement purification steps thereby provides a way to avoid an exponential scaling of the resources with the distance. This scheme is known as the quantum repeater (see e.g. \cite{Br98,Du98,Repeater,Ha06,RepeaterExp1,RepeaterExp2,RepeaterExp3,RepeaterExp4,RepeaterExp5,RepeaterExp6,RepeaterExp7,RepeaterExp8}). As shown in Refs. \cite{Br98,Du98}, one can establish entangled pairs over arbitrary distances with only polynomial overhead in the physical resources (number of required short distance pairs, parallel channels, repeater stations and time) even if local control operations are noisy (and local memory errors can be controlled \cite{Ha06}). The error thresholds for noisy local operations are of the order of percent. 

\begin{figure}[ht]
\begin{center}
\includegraphics[width=0.65\textwidth]{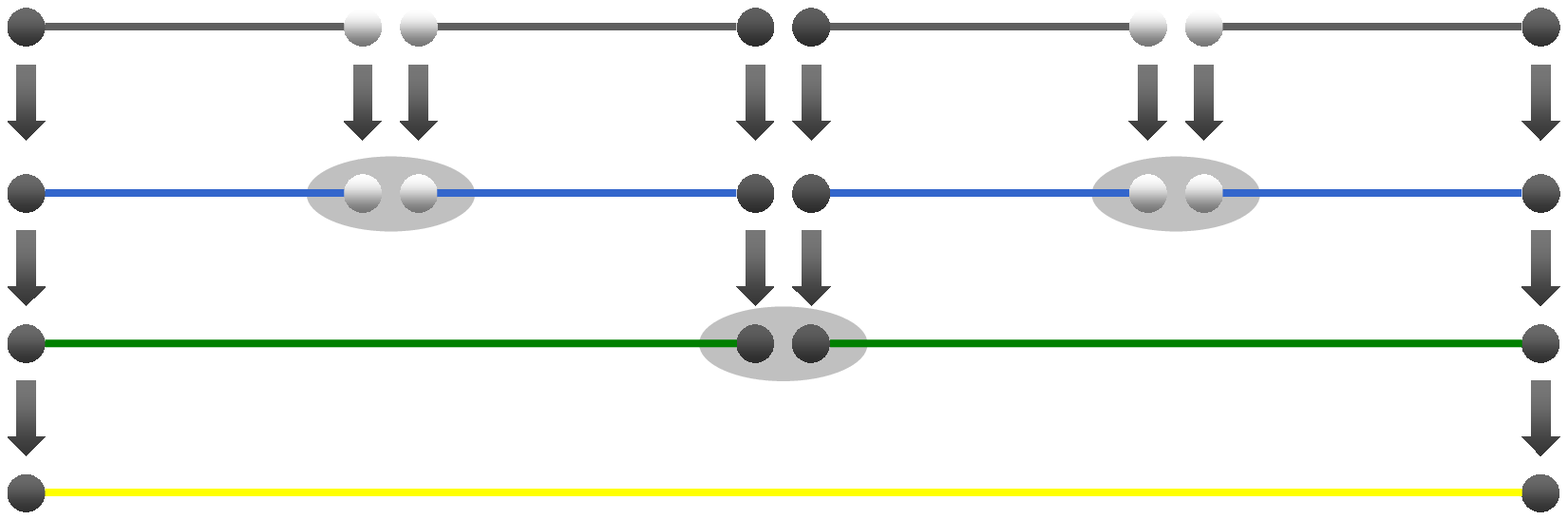}
\caption{Schematic sketch of the repeater setup. Connection and entanglement purification is used to generate large distance entanglement pairs with high fidelity.}
\label{Rep1}
\end{center}
\end{figure}

\begin{figure}[ht]
\begin{center}
\includegraphics[width=0.65\textwidth]{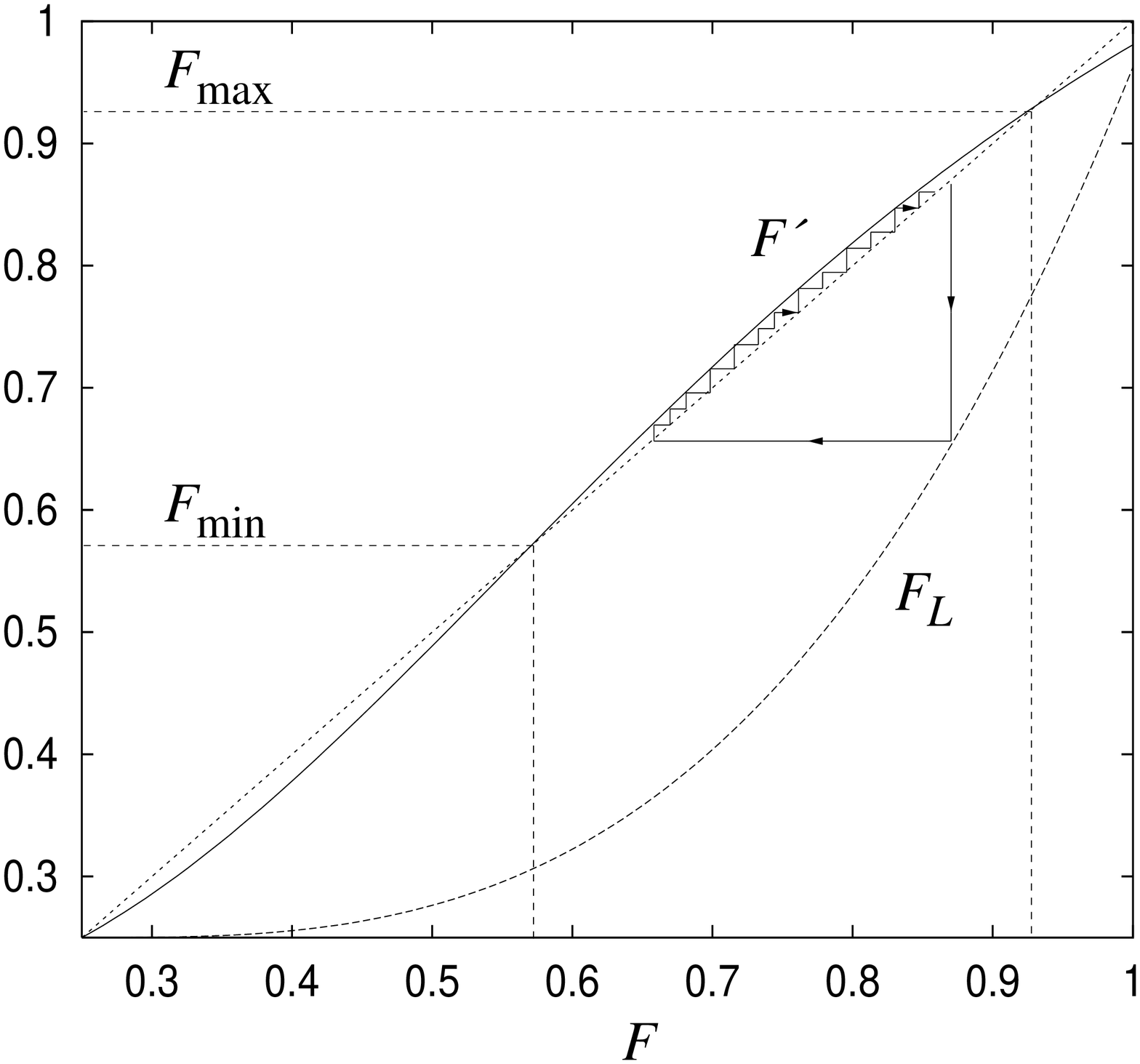}
\caption{Purification loop: Connection of $L$ elementary pairs and re--purification to initial fidelity $F$. In the text, $L=2$ is assumed. Figure taken from Ref. \cite{Br98}.}
\label{Repeater_figurethreshold}
\end{center}
\end{figure}

\begin{figure}[ht]
\begin{center}
\includegraphics[width=0.65\textwidth]{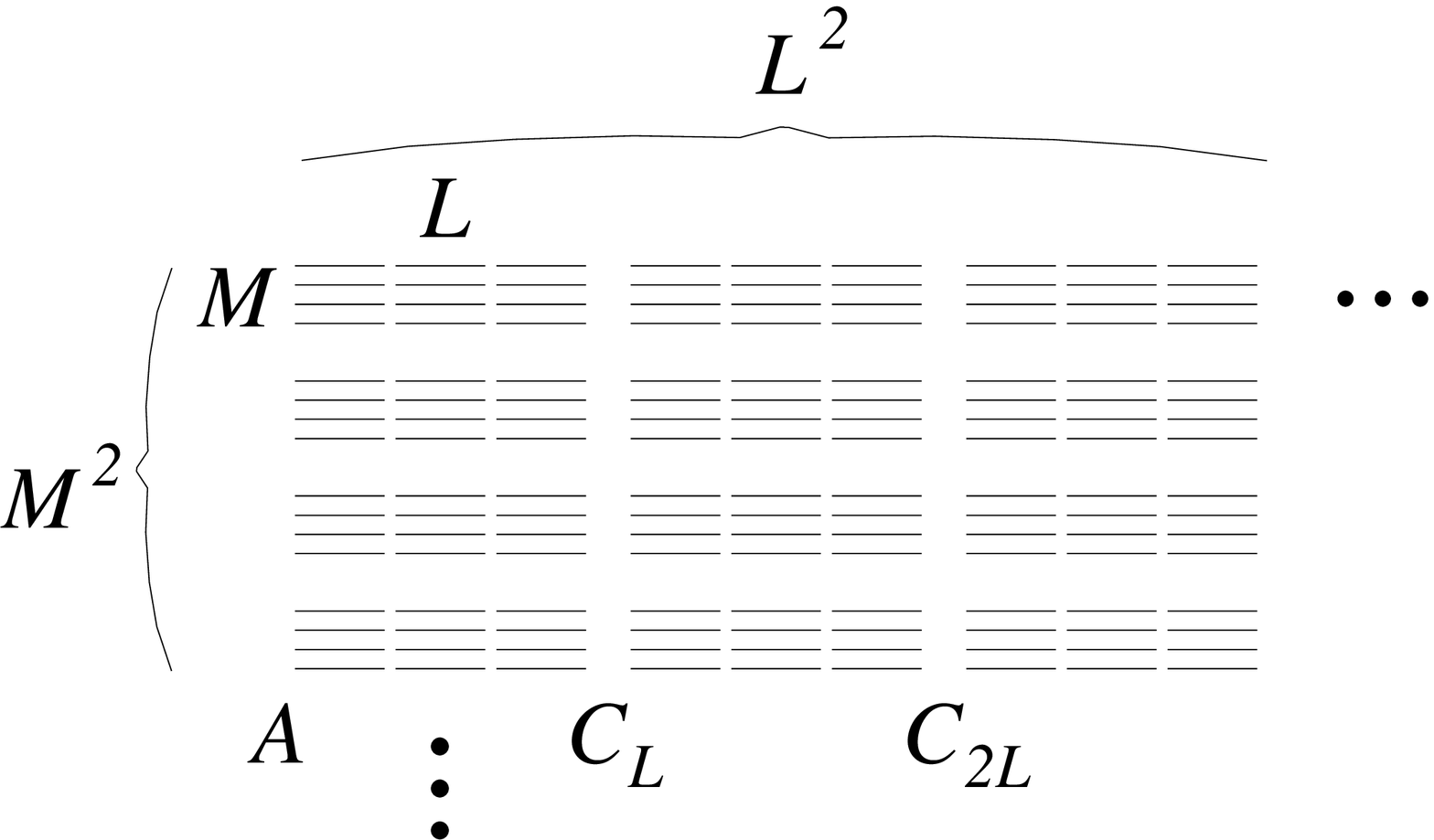}
\caption{Nested purification with an array of elementary EPR pairs. At each nesting level, $L$ elementary pairs are connected and re-purified using $M$ copies, where in the text $L=2$ is assumed. This yields to a polynomial scaling of the total elementary pairs with the distance. Figure taken from Ref. \cite{Br98}.}
     \label{Dur_Repeater_nested}
     \end{center}
\end{figure}

A brief sketch of the repeater protocol follows: For communication over a distance $N l_0$ we split the channel into $N$ segments of length $l_0$ and place repeater stations at distances $k l_0, k=0,1,\ldots ,N$. We assume that $l_0$ is sufficiently small such that the resulting state, when sending part of a maximally entangled state through the channel, is still distillable entangled and the absorption probability is sufficiently small. For optical fibers, distances of (several tens) of kilometers are reasonable. For simplicity, we assume that we have $2^n$ such segments, i.e. $N=2^n$. Several copies of short distance elementary pairs shared between all repeater stations are generated and purified to some working fidelity $F_0$. At repeater stations 2, 4, 6, etc. two adjacent elementary pairs are connected by performing entanglement swapping, thereby generating entangled pairs over distance $2 l_0$ with reduced fidelity shared between repeater stations $(1,3); (3,5); \ldots$. Purification of several copies of these pairs to the working fidelity $F_0$ results in a similar situation as previously, however the elementary pairs extend now over the distance $2 l_0$. Proceeding in this way, we obtain a nested scheme where the distance of the entangled pairs is {\em doubled} at each nesting level. At nesting level $n$, we obtain long distance entangled pairs of length $2^n l_0$ with fidelity $F_0$. It is straightforward to check that this scheme leads to a polynomial scaling of the resources with distance \cite{Br98}. The setup of the repeater is schematically sketched in Fig. \ref{Rep1}, while Fig. \ref{Repeater_figurethreshold} shows a elementary purification loop. The total number of required elementary short distance pairs are shown in Fig. \ref{Dur_Repeater_nested}.

Variations of the scheme based on the usage of different purification protocols --e.g. (nested) entanglement pumping rather than standard DEJMPS protocol-- have also been discussed and lead to a significant reduction of spatial resources (storage particles) to $n+1 \propto \log(N)$ at the cost of increased, although still polynomial, temporal resources \cite{Br98,Du98}. Further improvement to a {\em constant} overhead in spatial resources is possible (see Ref. \cite{Repeater}).

\subsection{Improving error thresholds in quantum computation}\label{EPPQC}

Under certain circumstances, entanglement purification can be used directly to weaken the requirements for fault--tolerant quantum computation \cite{Du03QC}. Consider a situation where $n$ systems, each of them possessing $d$ degrees of freedom, are available. For instance, one may think of $n$ neutral atoms or trapped ions, each of them constituting a $d$--level system. While typically only two of the levels are used for quantum computation, in principle more levels are available. In such a situation, one can show that the threshold for fault--tolerant quantum computation essentially only depends on the fidelity of single--system operations \cite{Du03QC}. Two--system operations, i.e. interactions between two systems, are typically more difficult to realize than single--system operations (e.g. operations on a single atom). However, it turns out that one can tolerate a noise level of more than $50\%$ for two--system operations, while still achieving fault tolerant quantum computation as long as the single system operations are of sufficiently high fidelity.

The basic idea is to use each $d$--level system to represent one qubit for computation, while the remaining degrees of freedom serve as auxiliary levels. The noisy two--system interaction serves to entangle auxiliary degrees of freedom, and one may use entanglement purification to increase the fidelity of this entanglement. Finally, high fidelity entangled states are used to realize two--system gates, e.g. by means of teleportation based gates \cite{Ci00,TBG1,TBG2,TBG3,TBG4,TBG5}. The fidelity of the two--system gate is essentially determined by the fidelity of the entangled state, which, in turn, is determined by the fidelity of single--system operations used in entanglement purification.

\begin{figure}[ht]
\begin{center}
\includegraphics[width=0.85\textwidth]{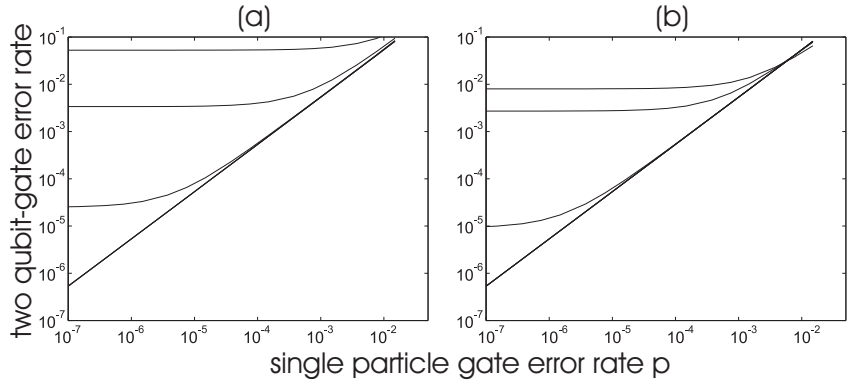}
\caption{Double logarithmic plot of achievable logical two--qubit gate error rate against single--particle error rate $p$ for fixed error rate of physical two--particle interaction of (a) $1.5*10^{-1}$, (b) $10^{-2}$. Curves from top to bottom correspond to no entanglement purification, entanglement pumping using 1, 2, 3 (or more) nesting levels respectively. Figure taken from Ref. \cite{Du03QC}.}
\label{FigQC1}
\end{center}
\end{figure}

We remark that at least four auxiliary levels should be available. By using nested entanglement pumping, as discussed in Sec. \ref{nestedEPP}, it turns out that for parameter regimes of practical importance, a few (2-3) nesting levels are sufficient to obtain high fidelity entanglement. This translates into a total requirement of about 16 levels per system, and a required error threshold of about $10^{-5}$ for single system operations to achieve errors of $10^{-4}$ for (logical) two--system operations, which is sufficient to achieve fault tolerant quantum computation. This is illustrated in Fig. \ref{FigQC1}. The error rate of the physical two--system operation can, however, be almost arbitrarily large (more than $50 \%$), and the two--system gates can even be probabilistic. 

A combination of (dynamical) decoherence free subspaces and entanglement purification is also conceivable. In Ref. \cite{Ta05PRL}, an entanglement purification protocol for spin degrees of freedom in electrically controlled semiconductor quantum dots has been put forward. The protocol is capable of purifying {\em encoded} states. The encoding is used to protect quantum information from dominant noise processes and is based on a dynamical decoherence free subspace. The purification protocol is capable to deal with other types of errors, where not only errors within the logical subspace, but also leakage errors can be handled. The purification protocol is constructed in such a way that it only makes use of resources available in such devices. One may also combine this encoded entanglement purification with a teleportation-based approach to implement non-local quantum gates, leading to an alternative way to obtain logical two-qubit gates for (fault-tolerant) quantum computation. Such a proposal for electrically controlled quantum dots has been put forward in Ref. \cite{Ta05Nature}.



\section{Entanglement purification protocols in multipartite systems}\label{multipartiteEPP}

We now turn to multipartite systems of $n$ spatially separated parties. We start by reviewing the concept of graph states, a family of $n$--qubit states of particular importance. Then we consider entanglement purification protocols for such graph states. The first protocol of this kind was introduced in Ref. \cite{Mu01} and further analyzed in Ref. \cite{Ma01}, and it is capable of distilling $n$--party GHZ states. Here, we will discuss recurrence and hashing protocols for all stabilizer states, or equivalently, all graph states. These protocols where introduced in Ref. \cite{Du03} and further elaborated in \cite{As04,Kr06}. 

\subsection{Graph states}\label{graphstates}

\begin{figure}[ht]
\begin{center}
\includegraphics[width=0.65\textwidth]{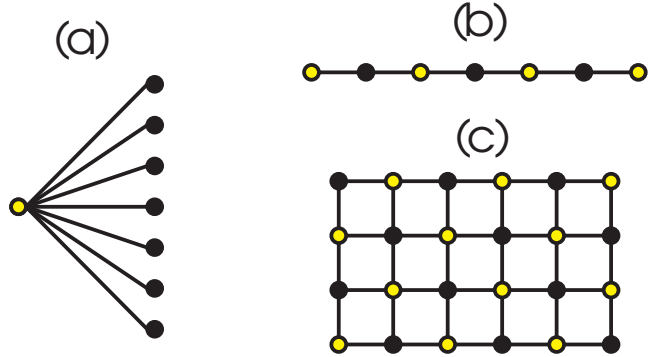}
\caption[]{\label{twoC} Examples of two--colorable graphs which correspond to (a) GHZ state; (b) linear cluster state; (c) two--dimensional cluster state. Vertices with same color are not connected by edges. Figure taken from Ref. \cite{Du05}.}
\end{center}
\end{figure}

We start by defining graph states. A graph $G=(V,E)$ is given by a set of $n$ vertices $V=\{1,2,\ldots ,n\}$ connected in a specific way by edges $E \in V^2$. To every such graph there corresponds a basis of $n$--qubit states $\{|\Phi_{\bm \mu}\rangle_G\}$, where each of the basis states $|\Phi_{\bm \mu}\rangle_G$ is the common eigenstate of $n$ commuting correlation operators $K_j^G$ with eigenvalues  $(-1)^{\mu_j}$, ${\bm \mu} = \mu_1\mu_2\ldots \mu_n$. To relax notation, we will sometimes omit the index $G$ and assume that an arbitrary but fixed graph $G$ is considered. Graph states fulfill the set of eigenvalue equations 
\begin{eqnarray}
K_j^G |\Phi_{\bm \mu}\rangle_G = (-1)^{\mu_j}|\Phi_{\bm \mu}\rangle_G,
\end{eqnarray} 
$j=1,\ldots,n$. The correlation operators are uniquely determined by the graph $G$ and are given by 
\begin{eqnarray}
K_j= \sigma_x^{(j)} \prod_{\{k,j\} \in E} \sigma_z^{(k)}.
\label{K}
\end{eqnarray}
A graph is called two--colorable if there exists two groups of vertices, $V_A$,$V_B$ such that there are no edges inside either of the groups, i.e. $\{k,l\} \not \in E$ if $k,l \in V_A$ or $k,l \in V_B$. For graph states associated with two--colorable graphs, which we call two--colorable graph states, we will split the multi--index ${\bm \mu}$ into two parts, ${\bm \mu}={\bm \mu}_A,{\bm \mu}_B$, belonging to subsets $V_A$ and $V_B$ respectively. See Fig. \ref{twoC} for examples of two-colorable graph states.

Graph states have first been introduced in Ref. \cite{Ra03}, generalizing the notion of cluster states as introduced in Ref. \cite{Br01}. For the related notion of graph codes see \cite{Sc01,Gra02}. A detailed investigation of their entanglement properties has been given in the paper by Hein et al. \cite{He03}, see also Ref. \cite{He06} for a recent review. Graph states occur in various contexts in quantum information theory, in which multi-party quantum correlations play a central role. Examples are multi-party quantum communication, measurement-based quantum computation, and quantum error correction. Prominent examples of two--colorable graph states are GHZ states, cluster states \cite{Br01} and codewords of error correction codes \cite{Sc01,Gra02} (see e.g. Ref. \cite{As04}). In fact, as shown in \cite{Ch04}, two colorable graph states are equivalent to codewords of CSS codes. We also remark that the correlation operators $\{K_j\}$ are the generators of the stabilizer group of the state $|\Phi_{\bm 0}\rangle_G$, and the corresponding description in terms of the stabilizers is also referred to as the stabilizer formalism \cite{Go97,Ho05}. 

We will also consider mixed states $\rho$, which for a given graph $G$ can be written in the corresponding graph state basis $\{|\Phi_{\bm \mu}\rangle_G\}$, 
\be
\rho =\sum_{{\bm \mu},{\bm \nu}} \lambda_{{\bm \mu}{\bm \nu}} |\Phi_{\bm \mu}\rangle\langle \Phi_{\bm \nu}|.
\ee
We will often be interested in fidelity of the mixed state, i.e. the overlap with some desired pure state, say $|\Phi_{\bm 0}\rangle_G$, $F= \langle \Phi_{\bm 0}| \rho |\Phi_{\bm 0}\rangle$. We remark that depolarization of $\rho$ to a standard form $\rho_G$,
\begin{eqnarray}
\rho_{G}=\sum_{\bm \mu} \lambda_{{\bm \mu}}|\Phi_{{\bm \mu}}\rangle\langle\Phi_{{\bm \mu}}|
\end{eqnarray}
can be achieved by randomly applying correlation operators $K_j$ \cite{Du03,As04} which is a multi--local operation. The diagonal elements, in particular the fidelity, are left unchanged by this depolarization procedure.
Note that both the notation and the description of the depolarization procedure are similar to the ones used for Bell states, which are in fact graph states with two vertices, connected by a single edge.


\subsection{Recurrence protocol for two--colorable graph states}\label{recurrenceMP} 

In the following, we will discuss a family of entanglement purification protocols that allow one to purify an arbitrary two--colorable graph state. To be precise, for each two colorable graph there exists a purification protocol which allows one to obtain the pure state $|\Phi_{\bm 0}\rangle_G$ as output state, provided the initial fidelity is sufficiently large. 
The recurrence scheme \cite{Du03,As04} for purifying a two--colorable graph state is very similar to the BBPSSW and DEJMPS protocol for purifying Bell pairs. We consider two sub--protocols, $P1$ and $P2$, each of which acts on two identical copies $\rho_1=\rho_2=\rho$, $\rho_{12}\equiv\rho_1\otimes \rho_2$. The basic idea consists again in transferring (non--local) information about the first pair to the second, and reveal this information by measurements. 

\begin{figure}
\begin{center}
\includegraphics[width=0.55\columnwidth]{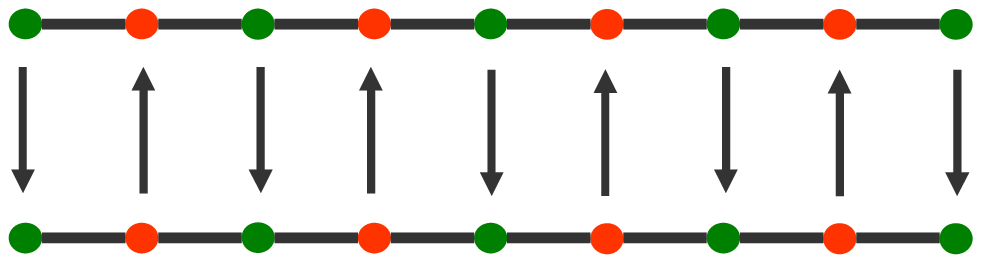}
\caption {Schematic representation of the multiparty entanglement purification for a 1D cluster state. }
\label{MEPPsetup}
\end{center}
\end{figure}

In sub--protocol $P1$, all parties who belong to the set $V_A$ apply local CNOT operations to their particles, with the particle belonging to $\rho_2$ as source, and $\rho_1$ as target (see Fig. \ref{MEPPsetup}). Similarly, all parties belonging to set $V_B$ apply local CNOT operations to their particles, but with the particle belonging to $\rho_1$ as source, and $\rho_2$ as target. The action of such a multilateral CNOT operation is given by \cite{Du03}
\begin{eqnarray}
|\Phi_{{\bm \mu}_{\bf A},{\bm \mu}_{\bf B}}\rangle|\Phi_{{\bm \nu}_{\bf A},{\bm \nu}_{\bf B}}\rangle\rightarrow|\Phi_{{\bm \mu}_{\bf A},{\bm \mu}_{\bf B}\oplus{\bm \nu}_{\bf B}}\rangle|\Phi_{{\bm \nu}_{\bf A}\oplus {\bm \mu}_{\bf A},{\bm \nu}_{\bf B}}\rangle
\label{psitopsi1}
\end{eqnarray}
where ${\bm \mu}_{\bf A}\oplus{\bm \nu}_{\bf A}$ denotes bitwise addition modulo 2. 

The second step of subprotocol $P1$ consists of a measurement of all particles of $\rho_2$, where the particles belonging to set $V_A$ [$V_B$] are measured in the eigenbasis $\{|0\rangle_x,|1\rangle_x\}$ of $\sigma_x$ [$\{|0\rangle_z,|1\rangle_z\}$ of $\sigma_z$] respectively. The measurements in sets $A$ [$B$] yield results $(-1)^{\xi_j}$ [$(-1)^{\zeta_k}$], with $\xi_j,\zeta_k \in\{0,1\}$. Only if the measurement outcomes fulfill the condition $(\xi_j+\sum_{\{k,j\}\in E}\zeta_k){\rm mod}2=0 ~\forall j$ ---which implies that the eigenvalues of all corresponding correlation operators $K_j$, $j\in V_A$ are +1, or equivalently ${\bm \mu}_{\bf A}\oplus{\bm \nu}_{\bf A}={\bf 0}$--- the first state is kept. In this case, one finds that the remaining state is again diagonal in the graph--state basis, with new coefficients
\begin{eqnarray}
\tilde\lambda_{{{\bm \gamma}_{\bf A}},{{\bm \gamma}_{\bf B}}} =\sum_{\{({{\bm \nu}}_{\bf B}, {{\bm \mu}}_{\bf B}) | {{\bm \nu}}_{\bf B} \oplus{{\bm \mu}}_{\bf B}={{\bm \gamma}_{\bf B}}\}} \frac{1}{2K}\lambda_{{{\bm \gamma}_{\bf A}},{{\bm \nu}_{\bf B}}}\lambda_{{{\bm \gamma}_{\bf A}},{{\bm \mu}_{\bf B}}},\label{mapP1}
\end{eqnarray}
where $K$ is a normalization constant such that ${\rm tr}(\tilde \rho)=1$, indicating the probability of success of the protocol.

In sub--protocol $P2$ the roles of sets $V_A$ and $V_B$ are exchanged. 
The action of the multilateral CNOT operation is in this case given by
\begin{eqnarray}
|\Phi_{{\bm \mu}_{\bf A},{\bm \mu}_{\bf B}}\rangle|\Phi_{{\bm \nu}_{\bf A},{\bm \nu}_{\bf B}}\rangle\rightarrow|\Phi_{{\bm \mu}_{\bf A}\oplus{\bm \nu}_{\bf A},{\bm \mu}_{\bf B}}\rangle|\Phi_{{\bm \nu}_{\bf A},{\bm \nu}_{\bf B}\oplus{\bm \mu}_{\bf B}}\rangle,
\label{psitopsi2}
\end{eqnarray} 
which leads to new coefficients
\begin{eqnarray}\tilde\lambda'_{{{\bm \gamma}_{\bf A}},{{\bm \gamma}_{\bf B}}} =\sum_{\{({{\bm \nu}}_{\bf A}, {{\bm \mu}}_{\bf A})| {{\bm \nu}}_{\bf A} \oplus{{\bm \mu}}_{\bf A}={{\bm \gamma}_{\bf A}}\}} \frac{1}{2K}\lambda_{{{\bm \nu}_{\bf A}},{{\bm \gamma}_{\bf B}}}\lambda_{{{\bm \mu}_{\bf A}},{{\bm \gamma}_{\bf B}}},\label{mapP2}
\end{eqnarray}
for the case in which the protocol $P2$ was successful.

The total purification protocol consists in a sequential application of sub--protocols $P1$ and $P2$. While sub--protocol $P1$ serves to gain information about ${\bm \mu}_{\bf A}$, sub--protocol $P2$ reveals information about ${\bm \mu}_{\bf B}$. Typically, sub--protocol $P1$ increases the weight of  all coefficients $\lambda_{{\bf 0},{\bm \mu}_{\bf B}}$, while $P2$ amplifies coefficients $\lambda_{{\bm \mu}_{\bf A},{\bf 0}}$. In total, this leads to the desired amplification of $\lambda_{{\bf 0},{\bf 0}}$.

The regime of purification in which these recurrence protocols can be successfully applied is rather difficult to determine analytically,  due to the non--trivial structure of the non--linear maps describing the protocol. Numerical investigation have been performed in Ref. \cite{As04}, and we refer the interested reader to this article for details. For special noise models, e.g. phase noise, the purification regime can be determined analytically, and provable optimal protocols with respect to purification range and yield can be found \cite{Ka06,Ka06b}. The mixed states considered under such a phase noise model are, in fact, thermal states of many--body spin hamiltonians defined via the corresponding graphs \cite{Ka06}.
We remark here that the fidelity does not provide a suitable measure to compare purification regimes for different number of particles $n$, as typically the required fidelity will decrease exponentially for all states. This is related to the exponential growth of the dimension of the Hilbert space with the number of particles $n$. One can alternatively consider the maximum acceptable amount of local noise per particle such that the state remains distillable by means of the recurrence protocol. That is, one assumes that each of the particles belonging to a given graph state is sent through a noisy quantum channel (e.g. a depolarizing channel) to its final location. One then finds for linear cluster states (or, more generally, all graph states with a constant degree) that the maximum acceptable amount of noise per particle is essentially independent of the particle number. For GHZ states, however, the acceptable amount of noise per particle decreases with increasing particle number. That is, GHZ states become more and more difficult to purify as the number of particles increases.

\subsubsection{Example: Binary--type mixture}

It is elucidating to consider the purification of a special family of states in some detail. We consider the example of mixed states of the form 
\begin{eqnarray}
\rho_{\cal A}\equiv \sum_{{\bm \mu}_{\bf A}} \lambda_{{\bm \mu}_{\bf A},{\bf 0}} |\Phi_{{\bm \mu}_{\bf A},{\bf 0}}\rangle \langle \Phi_{{\bm \mu}_{\bf A},{\bf 0}}|.
\label{rhoA}
\end{eqnarray} 
These states arise e.g. in a (hypothetical) scenario were all particles within set $V_A$ are only subjected to phase flip errors (described by $\sigma_z$), while all particles within set $V_B$ are subjected to bit flip errors ($\sigma_x$). The iterative application of protocol $P1$ is sufficient to purify states of the form Eq. (\ref{rhoA}), as only information about ${\bm \mu}_{\bf A}$ has to be extracted. A single application of protocol $P1$ leads again to a state of the form $\rho_{\cal A}$, with new coefficients 
\begin{eqnarray}
\tilde\lambda_{{\bm \mu}_{\bf A},{\bf 0}} = \lambda_{{\bm \mu}_{\bf A},{\bf 0}}^2/K,\label{binary1}
\end{eqnarray}
where $K=\sum_{{\bm \mu}_{\bf A}} \lambda_{{\bm \mu}_{\bf A},{\bf 0}}^2$ is a normalization constant indicating the probability of success of the protocol. That is, the largest coefficient is amplified with respect to the other ones. Iteration of the protocol $P1$ thus allows one to produce pure graph states $|\Phi_{{\bf 0},{\bf 0}}\rangle$ with arbitrary high fidelity, given the coefficient $\lambda_{{\bf 0},{\bf 0}}$ is larger than all other coefficients $\lambda_{{\bm \mu}_{\bf A},{\bf 0}}$. The family of states $\rho_{\cal A}$ includes states up to rank $2^{n_A}$, where $n_A$ denotes the number of particles in group $A$. Depending on the corresponding graph, $n_A$ can be as high as $n-1$ and hence the rank can be as high as $2^{n-1}$. 

As a concrete example, consider the one parameter family $\rho_{\cal A}(F)$ with $\lambda_{{\bf 0},{\bf 0}}=F$, $\lambda_{{\bm \mu}_{\bf A},{\bf 0}}=(1-F)/(2^{n_A}-1)$ for ${\bm \mu}_{\bf A} \not={\bf 0}$, where $F$ is the fidelity of the desired state. Application of protocol $P1$ maintains the structure of those states and leads to 
\begin{eqnarray}
\tilde F = \frac{F^2}{F^2+ (1-F)^2/(2^{n_A}-1)}.
\end{eqnarray}
This map has $\tilde F=1$ as attracting fixed point for all states with $F\geq 1/2^{n_A}$. The probability of success for a single step is given by $p=F^2+ (1-F)^2/(2^{n_A}-1)$. 

We also mention that other entanglement purification protocols for two-colorable graph states that makes use of error correction and error detection ideas have been introduced in Ref. \cite{Go06}. In particular, states corresponding to different graphs are used to purify a given noisy two-colorable graph state. The performance and robustness of these protocols with noisy gates is studied analytically, and it is shown that schemes with improved yield and scaling behavior can be obtained.

\subsection{Hashing protocol for two--colorable graphs states}\label{HashingMP}

In a similar way, one can design a hashing protocol for any two--colorable graph state. The first protocol of this type, capable of purifying GHZ states with non--zero yield, was introduced in Ref. \cite{Ma01}. Hashing protocols for arbitrary two--colorable graph states were presented in Refs. \cite{Ch04,As04}. The central tool in these protocols is already evident from Eqs. (\ref{psitopsi1}) and (\ref{psitopsi2}). These equations state how information about indices is transferred from one state to another. Information about all indices belonging to set $V_A$ is thereby transferred from copy one to copy two by the multilateral CNOT operations as specified in the first step of protocol $P1$, while information transfer occurs for all indices corresponding to set $V_B$ when the direction of CNOT operations is reversed (as it is done in $P2$). Again, by determining the parity of the bit values for random subsets ---which is done in a similar way as for Bell pairs, but here all bits belonging to set $V_A$ or set $V_B$ can be determined simultaneously---, one can learn the required information in such a way that the remaining ensemble is a tensor product of pure graph states (one needs to learn the classical information which non--local state is at hand). Notice that information transfer also takes place in the opposite direction, which is however not used.

The yield of the hashing protocol approaches unity for any state diagonal in the graph state basis with $\lambda_{{\bf 0}} \rightarrow 1$, independent of the specific form of the state. This implies that a given mixed state of sufficiently high fidelity $F$ can be purified with non--zero yield using the hashing protocol (combined with the depolarization procedure). Protocols with improved yield (by optimizing the information gain of the measurements) have been developed in Ref. \cite{Ho05a}.

\subsection{Recurrence protocol for all graphs states}\label{RecurrenceMP_all}

For the protocols described in the previous section, it is crucial that the underlying graphs of the states to be purified are two--colorable. It is, however, possible to obtain protocols that can purify {\em all} graph states. These protocols have been put forward in Ref. \cite{Kr06} and we briefly describe the basic idea here. Unlike in the case of two--colorable graph states, we have that the protocols do not operate on identical copies corresponding to the same graph, but on states corresponding to {\em different} graphs. 

Consider a $k$--colorable graph $G$, i.e. a graph where the set of vertices can be divided into $k$ non--connected subsets $\{V_1,V_2,\dots,V_k\}$. We denote the corresponding multi--indices by ${\bm \mu_1}, \ldots {\bm \mu_k}$. The purification of the associated graph state $|\Phi_{\bm 0}\rangle_G$ requires the alternating application of $k$ purification protocols. The $j^{\rm th}$ protocols serves to reveal information about ${{\bm \mu}_{j}}$, i.e. the indices associated to vertices in the set of qubits $V_j$. 

We will describe the $j^{th}$ purification protocol ${\mathcal P}_j$ in the following. The protocol consists essentially of two steps
\begin{itemize}
\item[(i)] Generation of a two--colorable graph state corresponding to graph $g_j$ from two copies of a graph state corresponding to graph $G$. 
\item[(ii)] Purification of one copy of a graph state corresponding to graph $G$ with help of the graph state corresponding to graph $g_j$.
\end{itemize}

\begin{figure}
\begin{center}
\setlength{\unitlength}{0.05625\columnwidth}
\begin{picture}(16,15)
\thinlines
\put(0,1.5){\includegraphics[width=0.9\columnwidth]{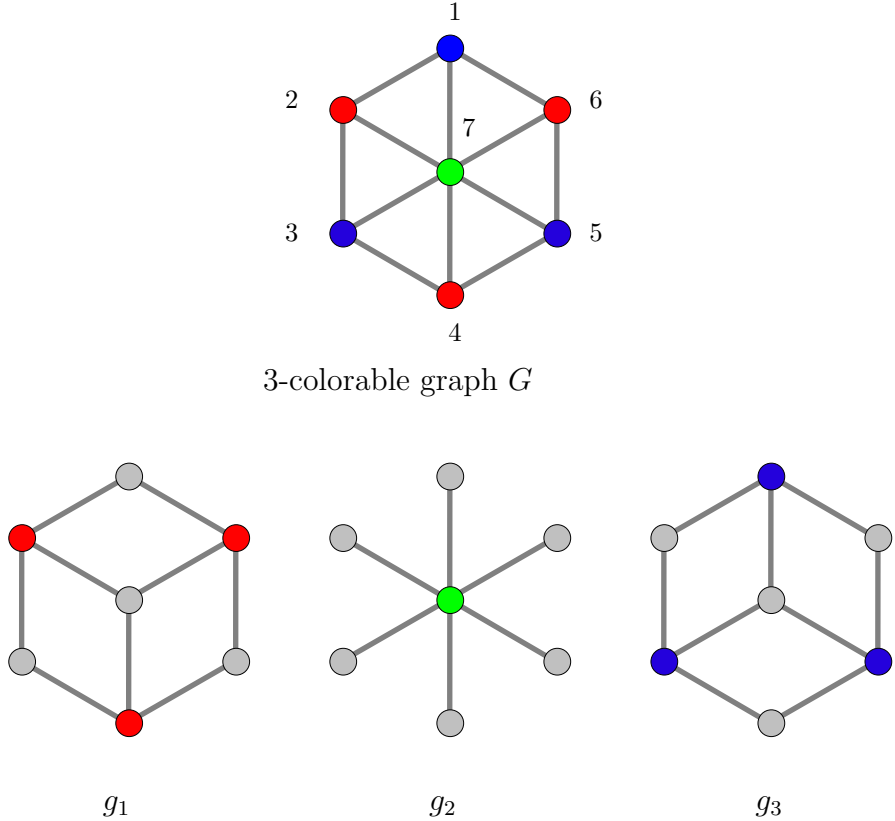}}
\put(5,10.5){3}
\put(5,12.9){2}
\put(7.95,14.5){1}
\put(10.5,12.9){6}
\put(10.5,10.5){5}
\put(7.95,8.7){4}
\put(8.2,12.4){7}
\put(4.6,7.8){\mbox{\large 3-colorable graph $G$}}
\put(1.7,0.2){\mbox{\large $g_1$}}
\put(7.6,0.2){\mbox{\large $g_2$}}
\put(13.5,0.2){\mbox{\large $g_3$}}
\end{picture}
\caption{A 3-colorable graph $G$ and the 3 corresponding two-colorable sub-graphs $g_1$, $g_2$, and $g_3$. $g_1$ corresponds to the red color (vertices 2,4 and 6), $g_2$ to the green color (vertex 7) and $g_3$ to the blue (vertices 1,3 and 5). Figure taken from Ref. \cite{Kr06}.
}
\label{fig:G and gj}
\end{center}
\end{figure}

Regarding (i), we define a two-colorable graph $g_j$ associated to $G$ as one which contains only the edges between the set $V_j$ and the remaining sets $\{V_i, i\not =j\}$, but where edges between the remaining sets are erased (see Fig. \ref{fig:G and gj} for an illustration). That is, the sets $\{V_i, i\not=j \}$ form a new set $V_{\bar j}=V\setminus V_j$. As shown in Ref. \cite{Kr06}, one can generate from two (noisy) copies of a graph state corresponding to graph $G$ a single (noisy) copy corresponding to graph $g_j$ by applying CNOT gates from the second to the first copy for all parties in the group $V_j$, and from the first copy to the second copy for all other parties, followed by $\sigma_z$ measurements performed on all particles of the second copy. Notice that CNOT gates between different copies do not only lead to a transfer of information, but also change the shape of the graph. In particular, a gate ${\tiny CNOT}_{a\to b}$ introduces new edges between the control qubit $a$ and all neighbors of the target qubit $b$ (or erases them if they are already there). In case of two--colorable graphs, the overall effect of the multilateral CNOT gates is that no additional edges are introduced (or more precisely they cancel each other). Here, the total effect of the CNOT operations is that all edges between groups $V_i$ and $V_l$ are erased whenever $i,l \not = j$ in the first copy, i.e. a two--colorable graph state corresponding to the graph $g_j$ is produced. 

An alternative to the $\sigma_z$ measurement of the second copy --which effectively simply erases the corresponding qubits-- is given by measurements in a different basis. More precisely, one may measure $\sigma_x$ on all qubits in group $V_j$, and $\sigma_z$ on all remaining qubits, and accepts the produced state only if the expectation value of the corresponding correlation operators $K_i$, $i \in V_j$ are +1. In this way, not only the shape of the graph is changed, but also a {\em purification} takes place as information about the first copy is revealed by the measurements of the second copy. 

In step (ii), we take two noisy states in such a way that the first state corresponds to graph $G$, $\rho_G$, and the second state to graph $g_j$,  $\rho_{g_j}$. The parties in the group $V_j$ apply CNOT gates from the second (control) state $\rho_{g_j}$ to the first 
(target) state $\rho_G$, while the remaining parties apply the CNOT gate in the opposite direction. The second state is measured in the eigenbasis of $\sigma_x$ for all parties in the group $V_j$ and in the eigenbasis of $\sigma_z$ for the remaining parties. The first state $\rho_G$ is kept only if all expectation values of the corresponding correlation operators $K_i$, $i \in V_j$ are +1. Again, the CNOT operation transfers information from the first state to the second, which is revealed by the measurements. The choice of different graphs for state 1 and 2 guarantees that the graph corresponding to the first copy remains unchanged. More precisely the action of CNOT operations is given by,
\be
|\Phi_{\bm \mu_{j}, \bm\mu_{{\bar j}}}\rangle_G|\Phi_{\bm \nu_{j}, \bm \nu_{{\bar j}}}\rangle_{g_j} \rightarrow |\Phi_{\bm \mu_{j},\bm \mu_{{\bar j}}\oplus \bm \nu_{{\bar j}}}\rangle_G|\Phi_{\bm \nu_{j} \oplus \bm\mu_{j} ,\bm \nu_{{\bar j}}}\rangle_{g_j},
\ee
which shows the transfer of information about the stabilizer eigenvalues between the two states. Notice the close similarity with Eq. (\ref{psitopsi2}), but also the crucial difference that here states corresponding to two different graphs are involved. After a successful purification step (i.e. obtaining proper measurement outcomes) one finds that the new matrix elements of $\rho'_G$ are given by
\be
{\lambda}'_{\bm{\gamma}_j,\bm{\gamma}_{\bar{j}}}=\frac{1}{\kappa}\sum_{\left\{(\bm{\mu}_{\bar{j}},\bm{\nu}_{\bar{j}})\mid \bm{\mu}_{\bar{j}}\oplus \bm{\nu}_{\bar{j}}=\bm{\gamma}_{\bar{j}}\right\}} \lambda_{\bm{\gamma}_j,\bm{\mu}_{\bar{j}}}\tilde{\lambda}_{\bm{\gamma}_j,\bm{\nu}_{\bar{j}}}
\ee
where the $\lambda$'s are the diagonal coefficients of the state $\rho_G$, and $\tilde \lambda$'s are the diagonal coefficients of the state $\rho_{g_j}$. As consequence, elements of the form $\lambda_{\bm{0},\bm{\gamma}_{\bar{j}}}$ are increased. One may say that purification takes place with respect to indices corresponding to parties in $V_j$.

The whole purification protocol consists of a sequential application of the sub-protocols ${\mathcal P}_j$ corresponding to all colors $j=1,\dots,k$. Even though there is a back-action of noise for the colors which are not purified for the step $j$, one obtains an overall increase of the fidelity $\lambda_{\bm{0}}$ if the fidelity of the initial state is sufficiently high. In fact, $\lambda_{\bm{0}}=1$ is an attractive fixed point of the protocol under the ideal local operations, which can be checked numerically.

\subsection{Purification of stabilizer states using stabilizer error correcting codes}

An alternative approach to purifying all stabilizer states has been taken in Ref. \cite{Gl06}. The multipartite entanglement purification protocols discussed there are $N \to M$ protocols, and stabilizing operators corresponding to error correction codes are measured locally on several copies of the stabilizer states to be purified. A link between the state to be purified and the code that can be used for purification is given. In particular, it is found that CSS states can be purified by CSS codes, while general stabilizer states can be purified by CSS-H codes.

\subsection{Breeding protocol for all graphs states}\label{HashingMP:all}

It is straightforward to construct a breeding protocol for all graph states using the ingredients presented in the previous section. This is also done in Ref. \cite{Kr06}. Several perfect copies of graph states corresponding to a graph $g_j$ are used to learn information about indices corresponding to qubits $V_j$ by means of a parity check, i.e. operating with CNOT gates on a (random) subset of copies with the perfect copy as a target. This is done for all groups $V_j$ independently, until complete knowledge about the ensemble of states at hand is obtained and hence $N$ copies of the pure graph state corresponding to graph $G$ are produced. Using the procedure described previously, graph states corresponding to graphs $g_j$ are generated and given back at the end (they where quasi borrowed to perform the breeding protocol). Two copies of the graph state $|\Phi_{\bm 0}\rangle_G$ are required to generate a single copy of $|\Phi_{\bm 0}\rangle_{g_j}$. Notice that also here an improved (adaptive) protocol is conceivable, which would lead to a higher yield. In addition, the generation of $n$ copies of $|\Phi_{\bm 0}\rangle_{g_j}$ from $m$ copies of $|\Phi_{\bm 0}\rangle_{G}$ might be possible with a rate $n/m$ larger than $1/2$, which would also increase the yield of the protocol.

A breeding protocol for all stabilizer states (inspired by the approach of Ref. \cite{Gl06}) with improved yield has recently been put forward in Ref. \cite{Ho06a}.

\subsection{Entanglement purification of non--stabilizer states}

While all bipartite and multipartite entanglement purification protocols we have described so far purify stabilizer states, i.e. state which are eigenstates of local stabilizer operators, a multipartite entanglement purification protocol was recently obtained \cite{My05} that allows one to purify a non--stabilizer state, in particular a $W$-state \cite{Du99W},
\begin{eqnarray}
|W\rangle=\frac{1}{3}(|001\rangle + |010\rangle + |100\rangle).
\end{eqnarray}
This protocol is a $3 \to 1$ protocol and stabilizing operators corresponding to an orthonormal basis including the W state are measured locally. 
Among other interesting features such as mutually unbiased bases, it has not only the 3--particle $W$ state but also maximally entangled states shared between two of the parties as attracting fixed points \cite{My05}. 

Furthermore, in Ref. \cite{Bo06}, a protocol for topological quantum distillation has been put forward. There, a new class of topological quantum error correction codes has been introduced, and it was shown how to perform arbitrary Clifford operations \cite{Go97} on encoded systems. As Clifford operations are the main ingredient of the purification protocols known so far, this provides a way of obtaining distillation protocols for topologically encoded systems.

\subsection{Multipartite recurrence protocols with noisy apparatus}

\begin{figure}[ht]
\begin{center}
\includegraphics[width=0.65\textwidth]{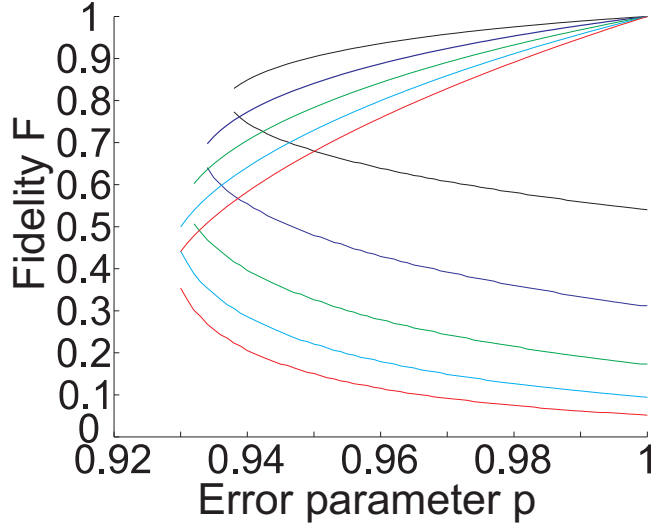}
\caption{Maximal reachable fidelity $F_{\rm max}$ and minimal required fidelity $F_{\rm min}$ plotted against error parameter $p$ (local operations) for density operators arising from single-qubit white noise. Curves from top to bottom (black, blue green, cyan, red) correspond to linear cluster states with $N=2,4,6,8,10$ particles. Figure taken from Ref. \cite{As04}.} 
\label{figurethreshold}
\end{center}
\end{figure}

Similar as for bipartite entanglement purification protocols, one can performed an analysis for multipartite entanglement purification protocols \cite{As04} with noisy apparatus. In this case, an intriguing question is the scalability of the process, i.e. whether e.g. the threshold for purification depends on the number of parties $N$. Numerical results for the purification range (minimal required and maximal reachable fidelity) as well as error threshold for linear cluster states of different size are given in Fig. \ref{figurethreshold}. Again, errors of the order of several percent are tolerable.

An important observation is that the threshold value $p_{\rm min}$ is for linear cluster states independent of the number of particles $n$. That is, also multipartite states of large number of particles can be successfully purified, and the requirements on local control operations are independent of the system size. This is not true when attempting to purify GHZ states \cite{As04}, where one finds that the required fidelity of local control operations depends on the particle number.  

The qualitative difference of cluster and GHZ states can already be understood from an analytically solvable toy model \cite{As04}, where one considers mixtures of GHZ states $|\Phi_{0,{\bm 0}}\rangle$ and $|\Phi_{1,{\bm 0}}\rangle$ and a restricted error model of only bit flip errors in set $V_B$, that preserve the structure of such states. Using the fact that bit flip errors in $V_B$ act like phase flip errors in $V_A$, and the fact that sub--protocol $P1$ is sufficient to purify such states, one obtains a lower bound on the threshold value $p_{\rm min}$ given by $p_{\rm min}= \left(\frac{1}{2}\right)^{1/(n-1)}$. This follows from arguments similar as in the derivation of purification curve for the bipartite BBPSSW protocol. Performing a similar analysis for binary--type mixtures of linear cluster states under this restricted noise model, one observes that the threshold value $p_{\rm min}$ is largely independent of the number of particles $n$, in agreement with the numerical observations for systems of up to size $n=10$ under a more general noise model. 





\section{Applications II}\label{applications2}

In this section we discuss  several applications of multiparty entanglement purification protocols. We understand that multipartite entangled states are {\em resources} to perform different tasks, ranging from multiparty secure applications such as secret sharing or Byzantine agreement to measurement--based quantum computation. Hence the generation of these entangled states with high fidelity is desirable, and here it is where multiparty entanglement purification comes into play.

\subsection{Quantum communication cost for multiparty state distribution}

\begin{figure}[ht]
\begin{center}
\includegraphics[width=0.85\textwidth]{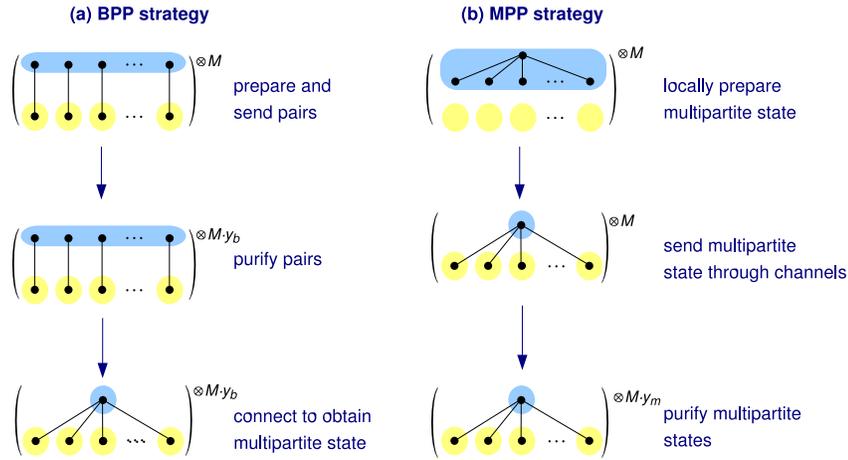}
\caption {Distribution of $N$-qubit GHZ states over noisy channels. (a) Bipartite entanglement purification strategy: Bell pairs are sent over the channels and purified using bipartite entanglement purification. The purified pairs are then connected to the desired GHZ state. (b) Multipartite entanglement purification strategy: Alice prepares the GHZ state locally and sends all but one of the particles through the channels. Then, the multiparty entanglement purification protocol is used. Figure taken from Ref. \cite{Kr05}.}
\label{QCC1}
\end{center}
\end{figure}

A scenario of particular interest is given when entangled states need to be distributed to spatially separated parties over noisy quantum channels in such a way that they are generated with high fidelity. As shown in Ref. \cite{Kr05}, there are several strategies conceivable to achieve this task. The two main strategies consist in (see Fig. \ref{QCC1})
\begin{itemize}
\item [(i)] the distribution of bipartite entangled pairs through noisy channels that are purified by using bipartite entanglement purification protocols; local generation of the desired multiparticle entangled state and distribution via teleportation to the different parties, using the purified entangled pairs.
\item [(ii)] the distribution of the locally generated multiparticle entangled states to the different parties through noisy quantum channels; the states are purified by using multiparticle entanglement purification protocols.
\end{itemize}
Apart from these two main strategies, a large number of intermediate strategies are conceivable, where parts of the multiparticle entangled state are distributed and purified, and these parts are connected later on to constitute the desired multiparticle entangled state, with possible intermediate or final purification steps. 

\begin{figure} [ht]
\begin{center}
\includegraphics[width=0.65\textwidth]{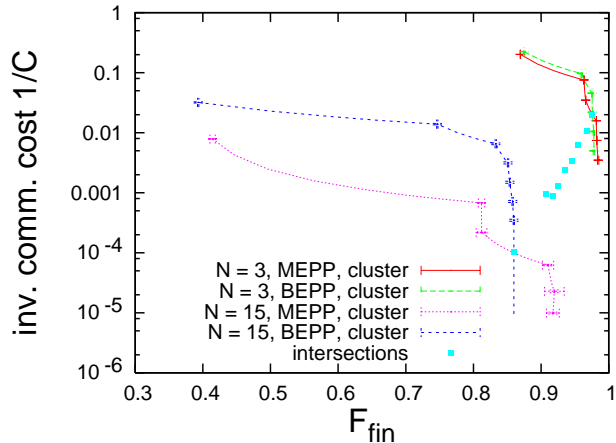}
\caption{(Color online.) Inverse of communication cost for different target fidelities for 3 (red solid line for multiparty entanglement purification and green dashed line for bipartite entanglement purification) and 15 (pink dotted line for multiparty entanglement purification and blue small dashed line for bipartite entanglement purification) qubit cluster states. The data points are the outputs for 1, 2, 3,~\dots iterations of the protocol. The intermediate points are obtained by mixing ensembles of different fidelities. For more than 6 steps, the difference between the reached fidelity and the maximum reachable fidelity is smaller than the uncertainty. For any number of parties, the curves representing the two strategies cross over. The disks give this cross-over for $N=3,4,5,6,7,8,9,10,15$. (That one curve seems to ``go back'' is just an artifact of the statistical inaccuracies of the Monte Carlo method.) Figure taken from Ref. \cite{Kr05}.} 
\label{QCC2}
\end{center}\bigskip
\end{figure}

\begin{figure}[ht]
\begin{center}
\includegraphics[width=0.65\textwidth]{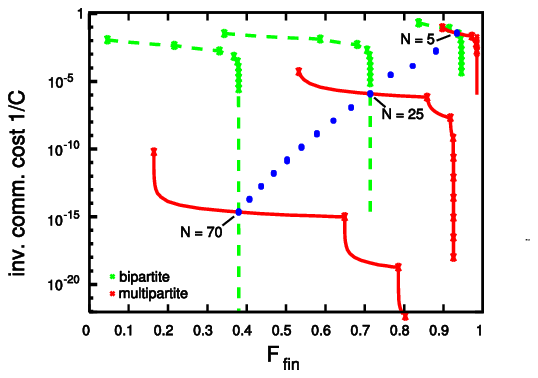}
\caption{Inverse of communication cost as function of final fidelity for a simplified noise model. Analytical calculation for GHZ states of different number of qubits $N$ varying from 5 to 70, with alteration probability for the channel and local noise of $(1-q)=0.1$ and $(1-q_l)=0.05$ respectively. The green dashed lines stand for multiparty entanglement purification strategy while the red solid lines stand for bipartite entanglement purification strategy. The blue circles give the crossing points for all number of parties between 5 and 70. Figure taken from Ref. \cite{Kr05}.}
\label{QCC3}
\end{center}
\end{figure}

A thorough analysis of the different strategies (see \cite{Kr05}) reveals that, in general, the multiparty strategy (ii) performs better if the desired target fidelity is high, while the bipartite strategy (i) may be used for lower target fidelities. In particular, the fact that multiparticle entanglement purification allows one to achieve higher fidelities than strategies based on bipartite purification makes multiparticle protocols the only choice if high target fidelities are required. The performance of the different strategies is measured by the quantum communication cost, that is, the total number of channel usages required to obtain on average a single copy of the desired multiparticle entangled state with sufficiently high fidelity. This is illustrated in Fig. \ref{QCC2} for linear cluster states and a generic noise model, while Fig. \ref{QCC3} shows results for GHZ state and a toy model for noise (see Ref. \cite{Kr05} for details).

\subsection{Secure state distribution}

The secure and secret distribution of an unknown multipartite state with high fidelity provides a basic quantum primitive, as multipartite entangled states can serve as resource to perform certain quantum information processing tasks. The specific type of entanglement determines the (class of) tasks that can be performed. Hence it is easy to imagine scenarios where the involved parties do not want any third party to learn which secret state they possess, and they wish at the same time their entanglement to be private. While in an idealized scenario where one assumes perfect local operations, this task can be achieved rather easily, under non--idealized conditions (as one typically faces) the problem becomes non--trivial. 
Multipartite entanglement purification is the main tool to achieve the secure and secret distribution of high--fidelity multipartite entanglement. However, standard entanglement purification protocols need to be modified to take care of additional secrecy and security requirements. In particular, even parties involved in the purification process may not be allowed to learn which state they are purifying.

\begin{figure}[ht]
\begin{center}
\includegraphics[width=0.85\textwidth]{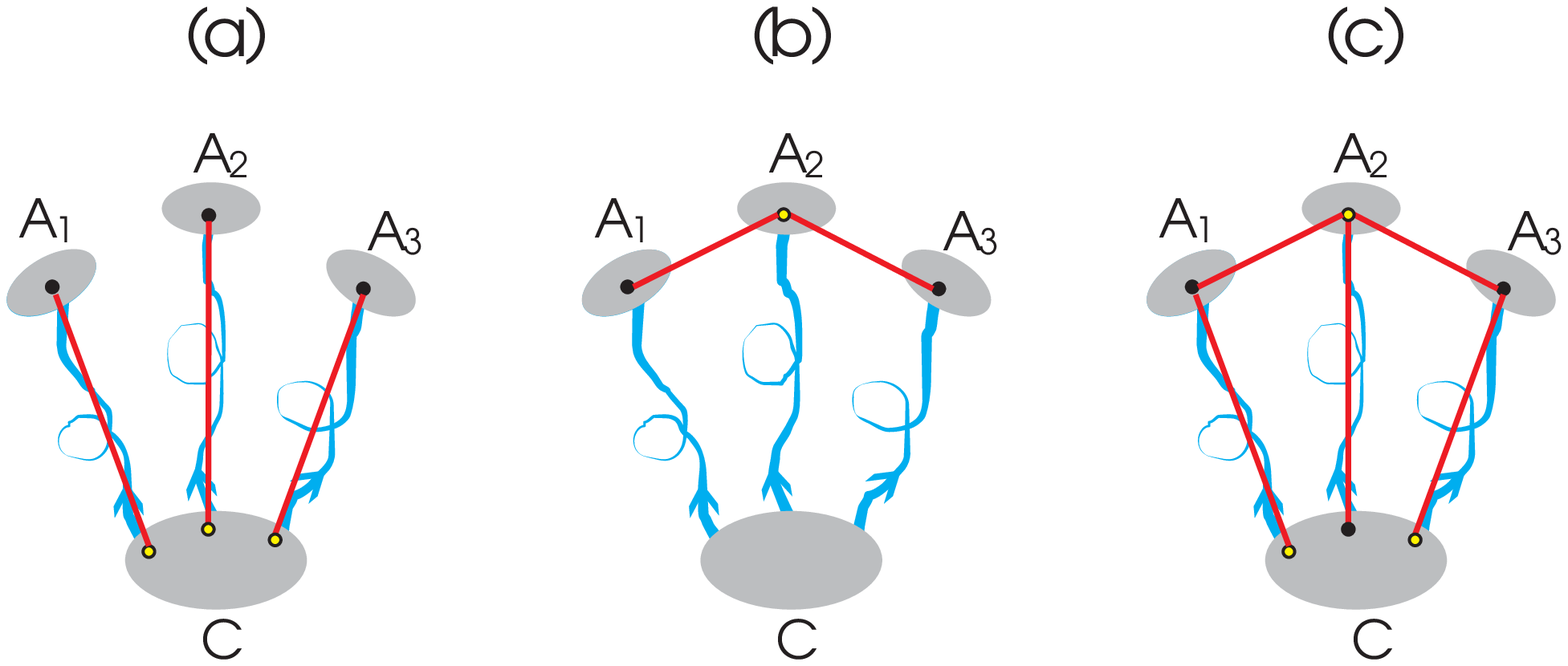}
\caption[]{Set--up for secure distribution of multipartite entangled states based on (a) Channel purification {\bf (i)}; (b) direct purification of multipartite entangled states {\bf (ii)}; (c) purification of enlarged entangled states {\bf (iii)}. Figure taken from Ref. \cite{Du05}.}
\label{SSDsetup}
\end{center}
\end{figure}

In Ref. \cite{Du05}, three different solutions to the secure--state distribution problem were put forward (see Fig. \ref{SSDsetup}). The first solution is based on bipartite entanglement purification, which serves to purify channels. Together with teleportation, this enables one to generate arbitrary multipartite entangled states. The second solution makes use of direct multipartite entanglement purification protocols, which is combined with basis randomization and adapted accordingly to ensure security. Security in the third solution, again based on direct multipartite purification, is ensured by purifying enlarged states. Each of the solutions offers its own advantages, and there exist in fact parameter regimes (for local noise, channel noise, desired target fidelity) such that one of the three schemes can be applied, while the other two fail. 

\subsection{Quantum error correction using graph states}

\begin{figure}[ht]
\begin{center}
\includegraphics[width=0.85\textwidth]{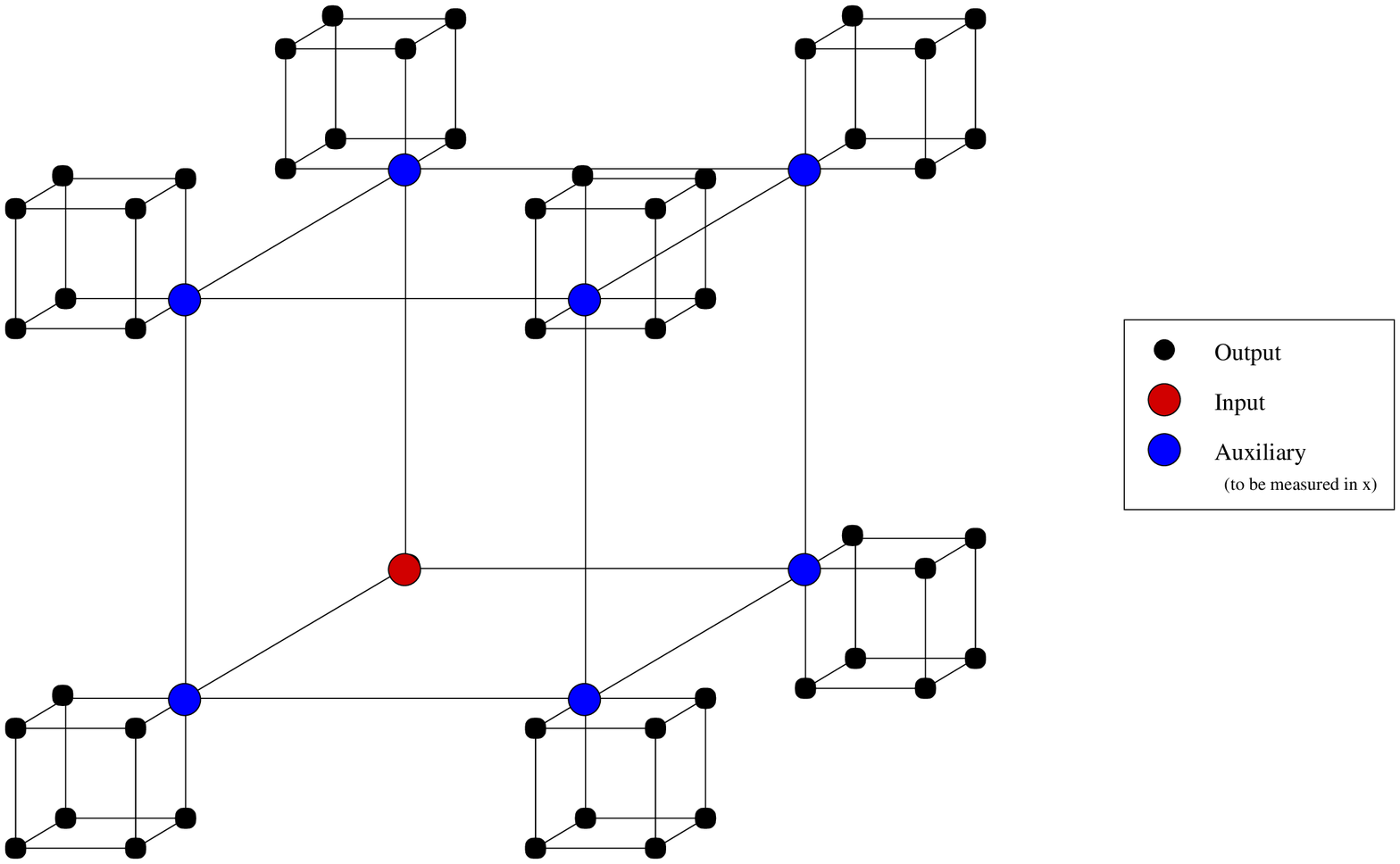}
\caption{Graph corresponding to a concatenated 7-qubit CSS code with input (red), auxiliary (blue) and output (black) vertices.
Figure taken from Ref. \cite{He06}.}
\label{CSScode}
\end{center}
\end{figure}

Since certain graph states constitute codewords of error correction codes, one may use the purification of these graph states to achieve high fidelity encoding without making use of complicated encoding networks \cite{As04}. In particular, the 7 qubit CSS code can be obtained by using a two--colorable graph state of eight vertices (a cube) as resource, and teleportation. Concatenated codes of this kind can be obtained by appending to each vertex of the cube another cube (see Fig. \ref{CSScode}). Encoding into the graph state can be achieved by a single Bell measurement \cite{As04}, where the qubit to be encoded is coupled by the Bell measurement to the $8^{\rm th}$ vertex of the cube. A similar procedure is considered for the five qubit code in Ref. \cite{Sc01}, where the notion of graph codes was introduced (see also \cite{Gra02}). The fidelity of the encoding mainly depends on the fidelity of the two--colorable graph state used in the procedure described above. Hence, multipartite entanglement purification can be applied to generate high fidelity entangled states which are then used to achieve high fidelity encoding. Notice that this can also be viewed as the purification of a quantum circuit, namely the encoding circuit. 

\subsection{Purification of circuits and one--way quantum computation}

In the one--way quantum computer model, a multipartite entangled state, the 2D cluster state \cite{Br01}, serves as universal resource for quantum computation \cite{Ra01,Ra03}. That is, given a cluster state of suitable size, an arbitrary quantum algorithm can be implemented by a sequence of single--qubit measurements. In a similar way, other graph states represent algorithm--specific resources, i.e. they allow one to implement a specific algorithm (depending on the graph state) by means of single qubit measurements \cite{Ra03}. In the presence of imperfect operations, the cluster-- or graph state may not be available with unit fidelity. However, entanglement purification may be applied to increase the fidelity and hence to reduce errors in quantum computation. To what extent the purification of graph states can be used in fault tolerant quantum computation is subject of current research. An interesting approach in this direction is the usage of {\em encoded} 2D cluster states \cite{Rau06}. The encoding serves to perform error correction and to allow for a fault--tolerant implementation of measurement based quantum computation. Also such encoded resource states may be purified using known entanglement purification protocols. Fault-tolerant one-way quantum computation with highly verified logical cluster states has recently been discussed in Ref. \cite{Fu06}.

We remark that other resource states have been identified recently to constitute a universal resource for measurement--based quantum computation \cite{Va06}. In particular, graph states corresponding to Triagonal--, Kogome- or Hexagonal lattices are also universal. The latter is of particular importance, as the corresponding states are less sensitive to local noise in the sense that their lifetime of entanglement is longer \cite{Du04}, and the multiparticle entanglement purification protocols that allow one to purify these states have less stringent error thresholds for local noise. This can be understood from the fact that the local degree of the graph determines (or at least strongly influences) the thresholds, as local noise only affects a given qubit and its neighbors in the graph. The degree of the hexagonal lattice is three, the minimal possible to obtain a universal resource for translationally symmetric lattice--type graphs \cite{Va06}. Notice that the 2D cluster state has degree four.

\subsection{Ancilla factory approach to fault--tolerant quantum computation}

Also in the network model for quantum computation, entanglement purification can be useful as already demonstrated in Sec. \ref{EPPQC}. Here we discuss a second possible application, making use of multiparty entanglement purification. In fault--tolerant quantum computation, quantum information is processed in an encoded form in a fault--tolerant way. The encoding into a higher--dimensional Hilbert space allows for the detection and correction of errors. To perform the error correction in a fault--tolerant way, one possible approach (see e.g. Refs. \cite{Kn04,St03,Br05,Rei05}) is to generate encoded ancilla particles in a predefined state $|0_L\rangle$, and use them for error syndrome extraction. In addition, logical single and two qubit gates can be performed in via gate--teleportation \cite{Ci00,TBG1,TBG2,TBG3,TBG4,TBG5}, i.e. by generating certain encoded entangled states which are used to implement gates on the logical qubits. 
When using (concatenated) CSS codes, the logical state $|0_L\rangle$ actually corresponds to a certain stabilizer (or graph) state, a highly entangled multiparticle state. Also the logical entangled states required for the teleportation--based gates are (highly entangled) multiparticle stabilizer states. The high--fidelity generation of these states is essential in the scheme, and in this context multiparticle entanglement purification, together with the usage of error detection schemes plays an important role. 

Recently Knill \cite{Kn04} has presented such an ``ancilla factory approach'' making use of ideas of resource state purification to estimate error thresholds for fault--tolerant quantum computation. The estimates he finds are of the order of $10^{-2}$, i.e. tolerable errors in operation are of order of one percent. This gives essentially the same order of magnitude as for multiparticle entanglement purification protocols. However, the approach still has a formidable overhead in resources, in particular for the high fidelity preparation of the encoded ancilla particles where billions of copies are required.


\section{Summary and outlook}\label{outlook}

In this article, we have briefly reviewed quantum error correction and illustrated the basic concepts of entanglement purification. We discussed the close relation between these two approaches in the context of quantum communication. We then concentrated on entanglement purification and illustrated the basic idea behind bipartite and multipartite entanglement purification protocols. We discussed several bipartite and multipartite entanglement purification protocols and their applications in quantum error correction, long distance quantum communication, multipartite secure applications, quantum error correction, gate purification and (fault tolerant) quantum computation. The remarkable robustness of entanglement purification protocols under the influence of errors in local control operations is central in this context, and establishes entanglement purification as a fundamental tool in quantum information procession. In particular, we have illustrated that entanglement purification is not only restricted to applications in bipartite quantum communication for which it was initially introduced, but can also be used for many other purposes. It is remarkable that entanglement purification remains a hot topic in the field of quantum information processing, and significant progress has been achieved in the last few years.

Despite of this progress, a number of open problems remain. For instance, we do not yet clearly understand the potential power and limitations of entanglement purification. The influence of noise and hence the corresponding error thresholds seem to be more relaxed than for general fault tolerant quantum computation. This appears to be related to the fact that two--way classical communication --as can be used in entanglement purification-- is provable stronger than one--way classical communication, and that one is attempting to protect known information (in the form of a maximally entangled state) rather than unknown information (as in the case of quantum error correction). Despite of this, as we have indicated in this article, indirect methods to use entanglement purification in the context of error correction or fault-tolerant quantum computation exist. These applications look very promising, but a more detailed analysis including a comparison of the resulting error thresholds needs to be performed. 

Other important open problems include the fundamental issue of what types of states can actually be purified by some genuinely multipartite entanglement purification protocol. Until recently, it appeared that this class is given by the set of stabilizer states. However, the discovery of a purification protocol for three qubit W--states shows that entanglement purification is not limited to stabilizer states, so the quest to identify and characterize the family of states where entanglement purification is possible remains open. In this context, it would e.g. be interesting to see whether protocols to purify W--states of more than 3 qubits can be found, whether these protocols can be generalized to purify all Dicke (or symmetric) states, or to discover new protocols for other kinds of maximally ---or perhaps even non--maximally-- multipartite entangled states. 

Another important, unresolved issue is the purification range of entanglement purification protocols, i.e. the identification of the set of mixed states that can be purified by a given entanglement purification protocol. In most cases only sufficient conditions for purification can be obtained via numerical Simulations, however an analytic treatment, in particular identifying necessary and sufficient purification conditions for a given protocol would be desirable. 

The construction of provable optimal entanglement purification protocols is another important challenge. Optimality can thereby be understood either with respect to the yield of the protocol, or with respect to the purification range. Despite considerable effort, such optimal entanglement purification protocols (with either optimality requirement) are known only for a few specific situations so far. 

Of particular importance in this context --but also extremely challenging-- is the extension of such investigations to the case of noisy local control operations. 
Given the possible practical applications of entanglement purification, most notable in the context of quantum repeaters for long distance quantum communication, but also for fault--tolerant quantum computation or quantum simulation \cite{Du07}, such optimizations are crucial to minimize the required physical resources and to simplify a practical realization.
More generally, finding new applications of entanglement purification in the broad context of quantum information theory and beyond remains an interesting and challenging task for future investigations.


\section*{Acknowledgements}
We especially thank A. Miyake and E. Hostens for reading the manuscript and providing valuable comments, and H. Aschauer for allowing us to use a figure from his PhD thesis. In addition we thank S. Anders, J. Calsamiglia, J. I. Cirac, A. Kay, C. Kruszynska, J. Pachos, J. Taylor and P. Zoller for fruitful collaborations on topics related to entanglement purification. This work was supported in part by the Austrian Science Foundation (FWF), the Deutsche Forschungsgemeinschaft (DFG), the European Union (PROSECCO)(QICS)(OLAQUI)(SCALA), and the Austrian Academy of Sciences (\"OAW) through project APART (W.D.).



\end{document}